 \documentclass[review,preprint,3p,12pt,numbers,sort&compress]{elsarticle}

\usepackage{amssymb}
\usepackage{amsmath}
\usepackage{graphicx}
\usepackage{subcaption} 
\usepackage{booktabs}
\usepackage{tabularx}
\usepackage{array}
\usepackage{xcolor}
\usepackage{hyperref}
\usepackage{mhchem}
\newcolumntype{C}[1]{>{\centering\arraybackslash}p{#1}}

\usepackage{lineno}

\journal{Combustion and Flame}

\begin{document}

\begin{frontmatter}



\title{Large eddy simulation of turbulent swirl-stabilized flames using the front propagation formulation: impact of the resolved flame thickness}



\author[1,2,3]{Ruochen Guo} 
\author[1]{Yunde Su\corref{cor1}} 
\ead{yundesu@scu.edu.cn}
\author[2,3]{Yuewen Jiang}

\cortext[cor1]{Corresponding authour}

\affiliation[1]{organization={National Key Laboratory of Fundamental Algorithms and Models for Engineering Numerical Simulation, Sichuan University},
             city={Chengdu},
             postcode={610207},
             state={Sichuan},
             country={China}}

\affiliation[2]{organization={University of Chinese Academy of Sciences},
             city={Beijing},
             postcode={100190},
             country={China}}
             
\affiliation[3]{organization={Institute of Engineering Thermophysics, Chinese Academy of Sciences},
             city={Beijing},
             postcode={100190},
             country={China}}


\begin{abstract}
This work extends the front propagation formulation\,(FPF) combustion model to large eddy simulation\,(LES) of swirl-stabilized turbulent premixed flames and investigates the effects of resolved flame thickness on the predicted flame dynamics.
The FPF method is designed to mitigate the spurious propagation of under-resolved flames while preserving the reaction characteristics of filtered flame fronts.
In this study, the model is extended to account for non-adiabatic effects and is coupled with an improved sub-filter flame speed estimation that resolves the inconsistency arising from heat-release effects on local sub-filter turbulence.
The performance of the extended FPF method is validated by LES of the TECFLAM swirl-stabilized burner, where the results agree well with experimental measurements.
The simulations reveal that the stretching of vortical structures in the outer shear layer leads to the formation of trapped flame pockets, which are identified as the physical mechanism responsible for the secondary temperature peaks observed in the experiment.
The prediction of this phenomenon is shown to be strongly dependent on the resolved flame thickness, when the filter size is used for modeling sub-filter flame wrinklings.
Without proper modeling of the chemical steepening effects, the thickness of the resolved flame brush is over-predicted, causing the flame consumption rate to be under-estimated.
Consequently, the flame brush detaches from the outer shear layer, resulting in a failure to capture the flame pockets and the associated secondary temperature peaks.
\end{abstract}


\end{frontmatter}


\section{Introduction}
\label{sec:Introduction}

Premixed swirl combustion is widely employed in modern gas turbines and aero-engines to achieve high combustion efficiency, low \ce{NOx} emissions, and \textcolor{black}{robust} flame stabilization~\cite{Gicquel2012,Candel2014,Huang2009,Aniello2023}.
However, significant multi-scale, unsteady, and non-linear interactions exist between the turbulent flow and the flame in these systems~\cite{O’Connor2023}.
These complexities pose substantial challenges for high-fidelity numerical simulations, which are essential for the design of next-generation clean propulsion systems.
Large eddy simulation\,(LES) has emerged as a promising approach for the high-fidelity simulation of turbulent combustion at affordable computational cost~\cite{Gicquel2012,Pitsch2006}.
It explicitly resolves the large-scale flame dynamics while modeling the small-scale flame-turbulence interactions.

In practical LES applications, however, the grid sizes are typically comparable to or even larger than the laminar flame thickness, due to limited computational resources.
When the grid scale is used as the filter size\,(i.e., in implicit LES), the filtered flame reaction zone is often under-resolved.
This leads to the flame resolution issue~\cite{Pitsch2006,Boger1998,Hawkes2000,Kuenne2017,VargasRuiz2025}, characterized by inaccurate predictions of the total reaction rate and spurious propagation of the filtered flame.  
Various strategies have been proposed to mitigate this issue.
For instance, the thickened flame LES\,(TFLES) method artificially broadens the filtered flame to resolve it on the coarse grid~\cite{Colin2000}.
Alternatively, spurious propagation can be avoided by tracking the flame surface as a geometric interface, as done in the G-equation model~\cite{Pitsch2005}, or alleviated using sub-filter interpolation techniques~\cite{Kuenne2017}.

To address the flame resolution issue while preserving the flame structure information, Kim and Su proposed the front propagation formulation\,(FPF) method for LES of turbulent premixed combustion~\cite{Kim2020}.
In this framework, the filtered chemical reaction rate is formulated as the product of the unburned density, the sub-filter flame consumption speed, and a regularized delta function.  
The introduction of the regularized delta function eliminates propagation errors inherent to under-resolved flame fronts. 
To reproduce the chemical steepening effects of premixed flames, the regularized delta function is modeled using the flame structure function, which depends on the flame parameters and the filter size.
While the FPF model has proven effective in LES of canonical configurations such as Bunsen and jet flames~\cite{Kim2020,Su2020,Su2025}, as well as spark-ignition engines~\cite{Su2021}, its application to complex swirl-stabilized combustion remains unexplored. 
In practical swirl-stabilized combustors, the strong coupling between the flame and the swirling vortical structures dictates flame stability and pollutant formation~\cite{Huang2009,Aniello2023,Wang2023,Chen2019}.
Moreover, such combustion processes typically involve fuel stratification and non-negligible heat loss effects\cite{Gicquel2012,Kraus2018,Benard2019,Xia2023,Zhao2025}.
Consequently, the applicability and performance of the FPF model in such complex swirling combustion configurations warrant further investigation.

\textcolor{black}{In LES of turbulent premixed combustion within the flamelet and thin reaction zones regimes, the resolved flame thickness represents the smallest scale of flame wrinklings that are explicitly captured, while smaller-scale wrinklings require explicit modeling.}
The magnitude of the resolved flame thickness is determined by the competition between turbulent transport, which tends to broaden the filtered flame, and chemical reactions, which induce thinning effects~\cite{Kim2020}.
In TFLES, by adjusting the thickening factor, the resolved flame thickness is artificially scaled to multiples of the laminar flame thickness, based on which the sub-filter flame wrinkling factor is defined~\cite{Colin2000,Veynante2015}.
In contrast, in other approaches such as the flame surface density model~\cite{Boger1998} or the unstrained/strained flamelet model~\cite{Langella2016}, the resolved flame thickness is not an explicit input parameter; instead, the filter size typically serves as the characteristic length scale for sub-filter modeling.
However, as discussed in~\cite{Kim2020}, when chemical steepening effects are not properly incorporated---especially in combination with gradient sub-filter turbulent transport, a common practice in combustion LES---the resolved flame thickness can significantly exceed the filter size.
This discrepancy between the resolved flame thickness and the filter size can degrade the fidelity of the LES.
Nevertheless, the impact of resolved flame thickness remains largely unexplored, particularly in LES of complex swirl-stabilized combustion.
\textcolor{black}{In the FPF method, the chemical steepening effect is governed by model parameters for the flame structure function.
While these parameters are theoretically linked to the filter size and laminar flame thickness,  their adjustability offers a unique opportunity to isolate and investigate the impact of the resolved flame thickness in LES.}

The objective of the present work is twofold.
First, we extend the FPF method to LES of turbulent swirling flames with fuel stratification and non-negligible heat loss.
The TECFLAM swirl burner, which has been the subject of many experimental~\cite{Keck2002,Schneider2005,Gregor2009} and numerical~\cite{Kuenne2017,Jones2012,Ayache2013,Butz2015} studies, is selected for this study.
The extended FPF method is validated \textit{a posteriori} through detailed comparisons between LES results and experimental measurements.
Second, we study the impact of the resolved flame thickness in LES of turbulent premixed swirling flames within the framework of FPF, with focus being placed on the flame dynamics and the interactions between resolved flames and vortical structures.

The remainder of this paper is organized as follows.
Details of the extended FPF framework, incorporating the heat-loss model and \textcolor{black}{an improved sub-filter flame speed evaluation method that resolves the inconsistency arising from heat-release effects on local sub-filter turbulence}, are presented in Section~\ref{sec:Model_formulations}.
The experimental configuration and numerical setup are summarized in Section~\ref{sec:Experimental_configuration_and_Numerical_Setup}.
The LES results of the TECFLAM swirl burner are discussed in Section~\ref{sec:Results}, focusing on the \textit{a posteriori} validation of the FPF model and the impact of the resolved flame thickness in LES. 
Finally, conclusions are drawn in Section~\ref{sec:Conclusions}. 

\section{Model formulations}
\label{sec:Model_formulations}
\subsection{Governing Equations}
\label{sec:Governing_Equations}

In partially premixed turbulent combustion under non-adiabatic conditions, the thermo-chemical state can be described using three scalars: the progress variable $c$, the mixture fraction $Z$, and the total enthalpy $h_t$, which is defined as the sum of the sensible enthalpy and the formation enthalpy.
$c$ is equal to 0 on the unburned side and 1 on the burned side.
In LES of partially premixed turbulent combustion, the flow and flame dynamics are governed by the filtered continuity equation, the filtered momentum equation, and the transport equations for the three Favre-filtered thermo-chemical scalars.  
These equations are expressed as,

\begin{equation}
\frac{\partial \bar{\rho}}{\partial t} + \frac{\partial \bar{\rho} \widetilde{u}_j}{\partial x_j} = 0,
\end{equation}
\begin{equation}
\frac{\partial \bar{\rho} \widetilde{u}_i}{\partial t} + \frac{\partial \bar{\rho} \widetilde{u}_j \widetilde{u}_i}{\partial x_j} = -\frac{\partial \bar{p}}{\partial x_i} + \frac{\partial \overline{\tau_{ij}}}{\partial x_j}
- \frac{\partial}{\partial x_j} \left[ \bar{\rho} \left( \widetilde{u_j u_i} - \widetilde{u}_j \widetilde{u}_i \right) \right], 
\end{equation}
\begin{equation}
\frac{\partial \bar{\rho} \tilde{c}}{\partial t} + \frac{\partial \bar{\rho} \widetilde{u}_j \tilde{c}}{\partial x_j}
= \frac{\partial}{\partial x_j} \left( \overline{ \rho D \frac{\partial c}{\partial x_j} } \right)
- \frac{\partial}{\partial x_j} \left[ \bar{\rho} \left( \widetilde{u_j c} - \widetilde{u}_j \tilde{c} \right) \right]
+ \bar{\rho} \widetilde{\omega}_c, 
\label{eq:transport_c}
\end{equation}
\begin{equation}
\frac{\partial \bar{\rho} \widetilde{Z}}{\partial t} + \frac{\partial \bar{\rho} \widetilde{u}_j \widetilde{Z}}{\partial x_j}
= \frac{\partial}{\partial x_j} \left( \overline{ \rho D \frac{\partial Z}{\partial x_j} } \right)
- \frac{\partial}{\partial x_j} \left[ \bar{\rho} \left( \widetilde{u_j Z} - \widetilde{u}_j \widetilde{Z} \right) \right] ,
\end{equation}
\begin{equation}
\frac{\partial \bar{\rho} \widetilde{h_t}}{\partial t} + \frac{\partial \bar{\rho} \widetilde{u}_j \widetilde{h_t}}{\partial x_j}
= \frac{\partial}{\partial x_j} \left( \overline{ \rho D \frac{\partial h_t}{\partial x_j} } \right)
- \frac{\partial}{\partial x_j} \left[ \bar{\rho} \left( \widetilde{u_j h_t} - \widetilde{u}_j \widetilde{h_t} \right) \right]. 
\end{equation}
Here, $\bar{(\cdot)}$ and $\tilde{(\cdot)}$ denote the filtering and Favre-filtering operation, respectively. 
$i$ and $j$ are the indexes of the spatial coordinate.
$\rho$ is the density, $u_i$ represents the velocity component, and $p$ is the pressure.
${\tau}_{ij}$ and $D$ denote the viscous stress tensor and the molecular diffusivity, respectively.
$\tau_{ij}$ is expressed as,
\begin{equation}
\tau_{ij}
=
-\frac{2}{3}\mu \frac{\partial u_j}{\partial x_j}\,\delta_{ij}
+
\mu\left(
\frac{\partial u_i}{\partial x_j}
+
\frac{\partial u_j}{\partial x_i}
\right),
\end{equation}
where $\delta_{ij}$ and $\mu$ represent the Kronecker delta and the molecular viscosity, respectively.
$\bar{\rho} \widetilde{\omega}_c$ is the filtered reaction source term to be modeled with a combustion model.
The filtered density $\bar{\rho}$ is evaluated from Favre-filtered progress variable $\tilde{c}$,
\begin{equation}
\frac{1}{\bar{\rho}} = \frac{\tilde{c}}{\rho_b} + \frac{1-\tilde{c}}{\rho_u},
\end{equation}
where $\rho_u$ and $\rho_b$ denote the densities of the unburned and burned mixtures, respectively. 
$\rho_u$ and $\rho_b$ are obtained from the equation of state as a function of the Favre-filtered mixture fraction $\tilde{Z}$ and total enthalpy $\tilde{h_t}$.
The molecular viscosity $\mu$ and diffusivity $D$ are estimated as  
$\mu = \mu_0 (\rho_0 / \bar{\rho})^{0.75}$ and 
$D = D_0 (\rho_0 / \bar{\rho})^{0.75}$, respectively,
where the subscript $0$ refers to the reference thermal state.

\subsection{Front propagation formulation model}
\label{sec:FPF_model}
In the FPF model~\cite{Kim2020}, the filtered chemical reaction source term on the right-hand side of Eq.~(\ref{eq:transport_c}) is formulated as,
\begin{equation}
\bar{\rho} \widetilde{\omega}_c = \rho_u S_{c,t}  \left| \nabla \psi(\tilde{c}) \right|,
\label{eq:source_FPF}
\end{equation}
where $S_{c,t}$ denotes the sub-filter flame consumption speed.
$\left| \nabla \psi(\tilde{c}) \right|$ is a regularized Dirac delta function with $\psi(1)-\psi(0)=1$.
The use of the regularized Dirac delta function in the FPF method guarantees that the integration of the reaction source term along the normal direction of the filtered flame front is independent of the resolution level, thus minimizing the spurious propagation of the filtered flame when it is under-resolved in LES.
The front structure function $\psi$ is designed to reproduce the mechanism of filtered flame propagation.
It is proposed as a function of $\tilde{c}$:
\begin{equation}
\psi = \alpha \tilde{c} + (1 - \alpha) \tilde{c}^{\gamma}
\label{eq:Psi},
\end{equation}
where model parameters $\alpha$ and $\gamma$ are estimated as,
\begin{equation}
\alpha = b_1 \left( \max\left( 0, \frac{\Delta}{l_F} - 2 \right) \right)^{1/2},
\label{eq:alpha}
\end{equation}
\begin{equation}
\gamma = \left( \gamma_0 - \gamma_\infty \right) \exp\left( -b_2 \frac{\Delta}{l_F} \right) + \gamma_\infty.
\label{eq:gamma}
\end{equation}
Here, $l_F$ is the laminar flame thickness, and $b_1$ and $b_2$ are model parameters, 
which have been determined from filtered DNS data and set to be $1/8$ and 1, respectively~\cite{Kim2020}.  
$\gamma_0$ is determined to reproduce the profile of the Arrhenius reaction in the reaction progress variable space. 
$\gamma_\infty$ is a constant set to be 1.5.

The proposed front structure function in the FPF method reproduces the limiting behaviors of the filtered flame fronts for both small and large filters.
The non-linear term with $\gamma>1$ ensures that the density-weighted displacement speed of iso-$\tilde{c}$ surfaces increases from the unburned to the burned side.
This property recovers the steepening of the resolved\,(filtered) flame brush by chemical reactions.
The thinning effects prevents the unbounded thickening of the filtered flame front, when the combined effects of molecular diffusion and modeled sub-filter scalar diffusion are gradient transport in practical LES.
More details on the FPF method, including the modeling of $\psi$ and comparisons with other premixed combustion models, can be found in a previous study~\cite{Kim2020}.

\subsection{Evaluation of sub-filter flame speed}
\label{sec:sub-filter_flame_speed}
With the flamelet assumption, the sub-filter flame consumption speed $S_{c,t}$ in Eq.~\eqref{eq:source_FPF} can be expressed as,
\begin{equation}
S_{c,t} = I_0 S_L \Xi_{\Delta}, 
\end{equation}
where $I_0$ denotes the stretch factor, $S_L$ is the laminar flame speed, and $\Xi_{\Delta}$ is the sub-filter wrinkling factor. 
$I_0$ can be set to $1$ when the stretch effects are negligible.
The laminar flame speed $S_L$ depends on the local equivalence ratio and the enthalpy defect.
The evaluation of $S_L$ under non-adiabatic conditions is described in Section~\ref{sec:heat_loss}.
The sub-filter wrinkling factor $\Xi_{\Delta}$ represents the ratio of the total flame surface area to the resolved flame surface area at a given filter size $\Delta$.
There have been several models for the closure of $\Xi_{\Delta}$.
In the present work, the Peters-Pitsch model~\cite{Pitsch2005,Peters1999} is chosen.
This model is formulated as,
\begin{equation}
\Xi_{\Delta} = \left[ 1 - \frac{c_3^2 \mu_t S_L}{2 c_1 \mu u'_{\Delta}} 
+ \sqrt{ \left( \frac{c_3^2 \mu_t S_L}{2 c_1 \mu u'_{\Delta}} \right)^2 
+ \frac{c_3^2 \mu_t}{\mu} } \right],
\label{eq:Xi_peters}
\end{equation}
where $\mu_t$ and $u'_{\Delta}$ denote sub-filter turbulent viscosity and sub-filter turbulent intensity, respectively. 
$c_1$ and $c_3$ are two model parameters, set to be 2.0 and 1.0, respectively. 

When the sub-filter turbulent parameters are used for modeling the sub-filter wrinkling factor, as is the case in Eq.~\eqref{eq:Xi_peters}, errors in the predicted flame propagation can appear due to the inconsistency between the model formulation and the practical usage.
In the derivation of the wrinkling factor models, the turbulent parameters in the models are typically defined as those in the incoming flow~\cite{Peters1999,You2020}.
On the other hand, in practical LES, $\Xi_{\Delta}$ is typically estimated using local sub-filter turbulent parameters which are strongly affected by the combustion heat release within the filtered flame brush~\cite{O'Brien2017,MacArt2021,Su2022,Wang2024} and thus differ from those in the unburned side.
This inconsistency causes the incorrect prediction of filtered flame propagation, when the filtered flame is resolved but not tracked.
To resolve this inconsistency, the wrinkling factor within the filtered flame brush is estimated by extending the values near the unburned side, where heat release is negligible, along the surface normal and imposing a zero-gradient condition.
Numerical tests have shown that this treatment improves the prediction of flame dynamics in LES of swirl-stabilized combustion.

\subsection{Heat loss modeling}
\label{sec:heat_loss}

Flame propagation is significantly influenced by both the mixing state and heat loss, the latter of which is inherent to practical swirl-stabilized combustion.
To account for heat loss effects in the FPF method, a non-adiabatic library for the laminar flame speed, $S_L$, and flame thickness, $l_F$, is constructed using simulations of one-dimensional freely propagating laminar premixed flames under non-adiabatic conditions with different equivalence ratios~\cite{Massey2021,Zhang2023a,Zhao2025}.
In these simulations, the energy equation is modified as,
\begin{equation}
\dot{m} C_p \frac{d T}{d x}
= \frac{d}{d x}\!\left( \lambda \frac{d T}{d x} \right)
- \sum_{k=1}^{K} C_{p,k} \, j_{k} \, \frac{d T}{d x}
- (1-\kappa_h)\sum_{k=1}^{K} h_k \dot{\omega}_k W_k .
\end{equation}
where $\dot{m}$, $C_p$, and $\lambda$, denote the mass flow, heat capacity, and thermal conductivity, respectively; $C_{p,k}$, $j_{k}$, $h_k$, $\dot{\omega}_k$ and $W_k$ are the heat capacity, diffusive mass flux, sensible enthalpy, reaction rate, and molecular weight of species $k$, respectively.
$\kappa_h$ is a heat loss factor used to quantify the extent of heat loss.
In the non-adiabatic library, the level of heat loss is parameterized using the total enthalpy, $h_t$.
Consequently, during the LES, the local values of $S_L$ and $l_F$ used in the combustion model are retrieved via interpolation based on the instantaneous local equivalence ratio and Favre-filtered total enthalpy $\widetilde{h_t}$.

\section{Experimental configuration and Numerical Setup}
\label{sec:Experimental_configuration_and_Numerical_Setup}

\subsection{Experimental configuration}
\label{sec:Experimental_configuration}

The target configuration is the unconfined TECFLAM swirl-stabilized burner, which has been extensively investigated both experimentally~\cite{Keck2002,Schneider2005,Gregor2009} and numerically~\cite{Kuenne2017,Jones2012,Ayache2013,Butz2015,Ren2024a}. 
In this configuration, a fully premixed methane--air mixture at $T = 300$\,K is injected through an annular slot surrounding a central bluff body into a co-flowing air stream. 
The bluff body has a radius of 15\,mm, and the annular gap width is also 15\,mm~\cite{Schneider2005}.
This study focuses on an operating condition with a Reynolds number of $Re = 10,000$\,(based on the nozzle exit bulk velocity and bluff body diameter) and a geometric swirl number of $0.75$. 
The mixture equivalence ratio is $0.833$, and the coflow velocity is $0.5$\,m/s. 
Regarding the thermal boundary condition, the bluff body is water-cooled internally to maintain a surface temperature of $353$\,K~\cite{Gregor2009,Butz2015}.
Under these operating conditions, the flame lies at the crossover between corrugated flamelets and thin reaction zones in the revised Borghi/Peters regime diagram~\cite{Schneider2005}.

Both non-reactive and reactive cases were performed in the experiment.
Under non-reactive conditions, the mean and fluctuating velocities in the axial, radial, and azimuthal directions at different axial locations downstream of the nozzle exit were measured~\cite{Schneider2005}.
For the reactive case, in addition to the velocity statistics, the mean and fluctuating values of the mixture fraction and temperature were also obtained~\cite{Schneider2005,Gregor2009}.

\subsection{Numerical setup}
\label{sec:Numerical_setup}
The governing equations for the LES are discretized using a finite difference scheme on a non-uniform staggered grid in cylindrical coordinates~\cite{Desjardins2008,Su2018}.  
The convective terms in the continuity and momentum equations are discretized using the second-order central energy-conservative scheme~\cite{Desjardins2008}, and the convective terms in the scalar transport equations and the source term in the FPF model are discretized using the fifth-order weighted essentially non-oscillatory\,(WENO) scheme~\cite{Jiang1996}.  
The combination of the second-order central scheme for mass conservation and the WENO scheme for scalar transport equations ensures the numerical stability under steep scalar gradients near the filtered flame front while maintaining consistency among different conservation laws~\cite{Su2018}.  
The viscous terms in the momentum equations and the diffusion terms in the scalar transport equations are discretized using the standard second-order central scheme.  
Time integration is performed using the second-order semi-implicit Crank--Nicolson scheme~\cite{Desjardins2008,Pierce2001}.

\begin{figure*}[!t]
	\centering
	\includegraphics[width=0.8\textwidth]{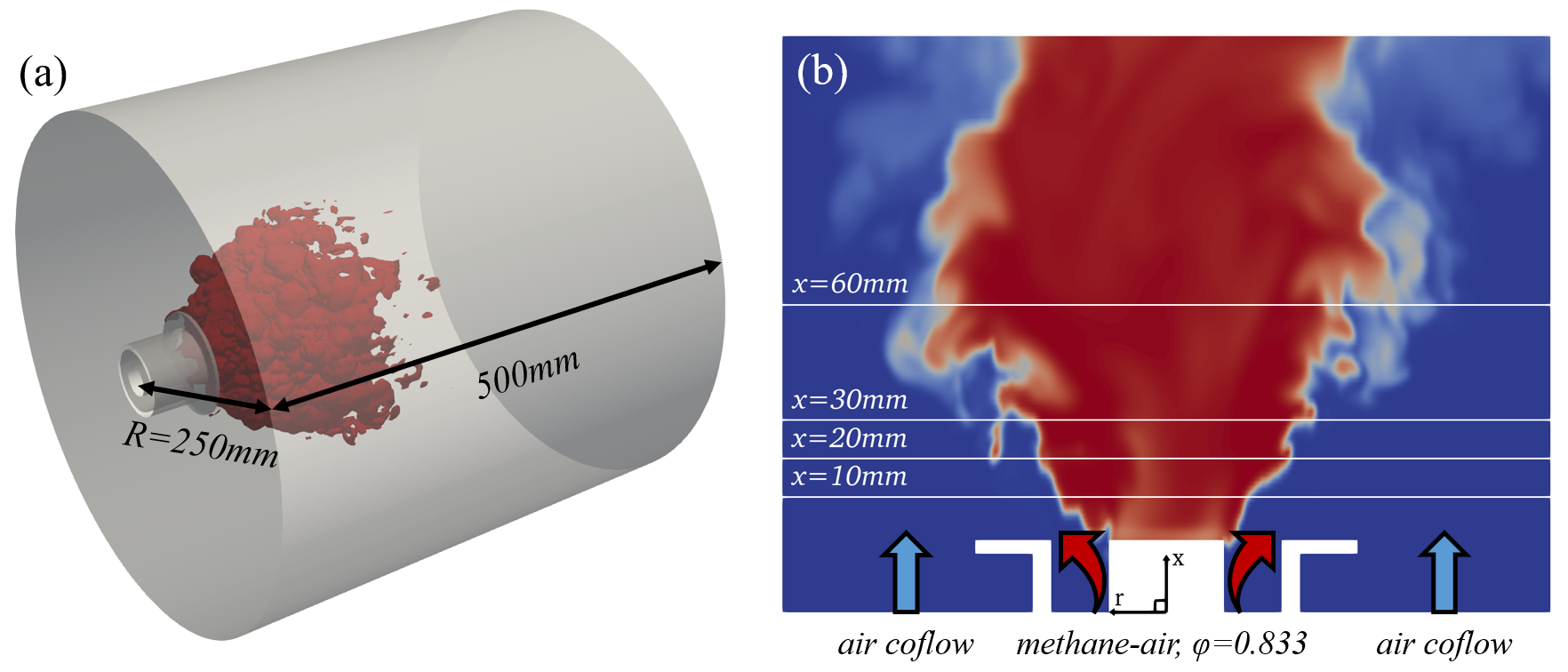}
	\caption{3-dimensional view\,(a) and radial section\,(b) of the computational domain.}
    \label{fig:domain_location}
\end{figure*}

The computational domain corresponds to a cylinder with a radius of 250\,mm and a height of 500\,mm, as illustrated in Fig.~\ref{fig:domain_location}.
The domain is discretized using a cylindrical grid with $N_x \times N_r \times N_{\theta} = 176 \times 176 \times 256$ points in the axial, radial, and azimuthal directions, respectively.
The grid is refined within the reactive region in the axial and radial coordinates, while remaining uniform in the circumferential direction.
A grid convergence analysis is provided in the Appendix, demonstrating that the current resolution is sufficient for the objectives of the present work.

In the simulations, a turbulent swirling inflow is injected into the computational domain through the annular slot inlet.
This inflow is generated using the linear forcing method~\cite{Zhang2023a,Pierce2001} with a prescribed swirl number of $0.75$.
A uniform coflow of $0.5$\,m/s is prescribed at the inflow boundary surrounding the slot inlet.
A convective boundary condition is applied at the domain outlet.
The surface temperature of the bluff body is set to $353$\,K to match the experimental measurements, while an adiabatic boundary condition is imposed on the remaining walls.

The sub-filter turbulent viscosity and diffusivity are modeled using the Lagrangian dynamic turbulence model~\cite{Meneveau1996}.  
The filtered reaction source term in the transport equation of the Favre-filtered progress variable is modeled using the extended FPF method, as detailed in Section~\ref{sec:FPF_model}.
The filter size used in the combustion model is set as the grid size.
The one-dimensional simulations for constructing the non-adiabatic library are performed using Cantera~\cite{Cantera} with the GRI–Mech~3.0 chemical mechanism~\cite{GRI-Mech_3.0}.
Both non-reactive and reactive cases are simulated in the present work.
The non-reactive simulation serves to validate the boundary conditions, the sub-filter turbulence model, and the numerical discretization.

In the reactive simulation, the FPF model parameters $\alpha$ and $\gamma$ are determined using Eqs.~\eqref{eq:alpha} and~\eqref{eq:gamma}, based on the filter size and flame parameters near the slot outlet.
This results in $\alpha = 0$ and $\gamma = 2.5$.
Additional sensitivity tests, where $\alpha$ and $\gamma$ are allowed to vary depending on the local filter size and mixture state, showed no noticeable differences in the results.
To investigate the impact of the resolved flame thickness on LES of turbulent swirling flames, a comparative simulation with $\alpha=0$ and $\gamma=1.0$ is also performed.
Under this condition, the thinning effects due to chemical reactions are absent in the LES.

The maximum Courant–Friedrichs–Lewy\,(CFL) number is maintained at approximately $0.5$ across the entire computational domain for all the simulations.
Post-processing statistics are collected over a duration of $0.65$\,s, after the statistically steady state is achieved.
This duration corresponds to approximately 6 flow-through times, where the flow-through time is defined as the ratio of the domain length to the bulk velocity of the slot inlet.

\section{Results and Discussions}
\label{sec:Results}
\subsection{Cold flow simulation}
\label{sec:Cold_flow_results}
Figures~\ref{fig:axial_reac},~\ref{fig:radial_reac}, and~\ref{fig:azimuthal_reac} compare the mean and root mean square\,(RMS) velocities in the axial, radial, and azimuthal directions, respectively, between the LES predictions and experimental results.
The simulation satisfactorily captures the magnitude and position of the peak mean velocity components, indicating that the radial expansion of the swirling flow downstream of the slot inlet is correctly reproduced. 
Furthermore, the negative axial velocities show good agreement with measurements, demonstrating that the central recirculation zone is well resolved. 
Regarding the turbulent statistics, the fluctuating velocity components are also predicted with reasonable accuracy across different axial positions. 
In particular, the dual-peak structures in the fluctuation profiles of the axial and azimuthal components, which correspond to the inner and outer shear layers, align closely with the experimental data. 
The overall agreement between the LES results and experimental measurements validates the suitability of the present numerical setup—including swirling inflow, numerical discretization, boundary conditions, and the sub-filter turbulence model—for simulating the flow field of the TECFLAM burner.

\begin{figure}[htb]
	\centering
	\includegraphics[width=0.8\textwidth]{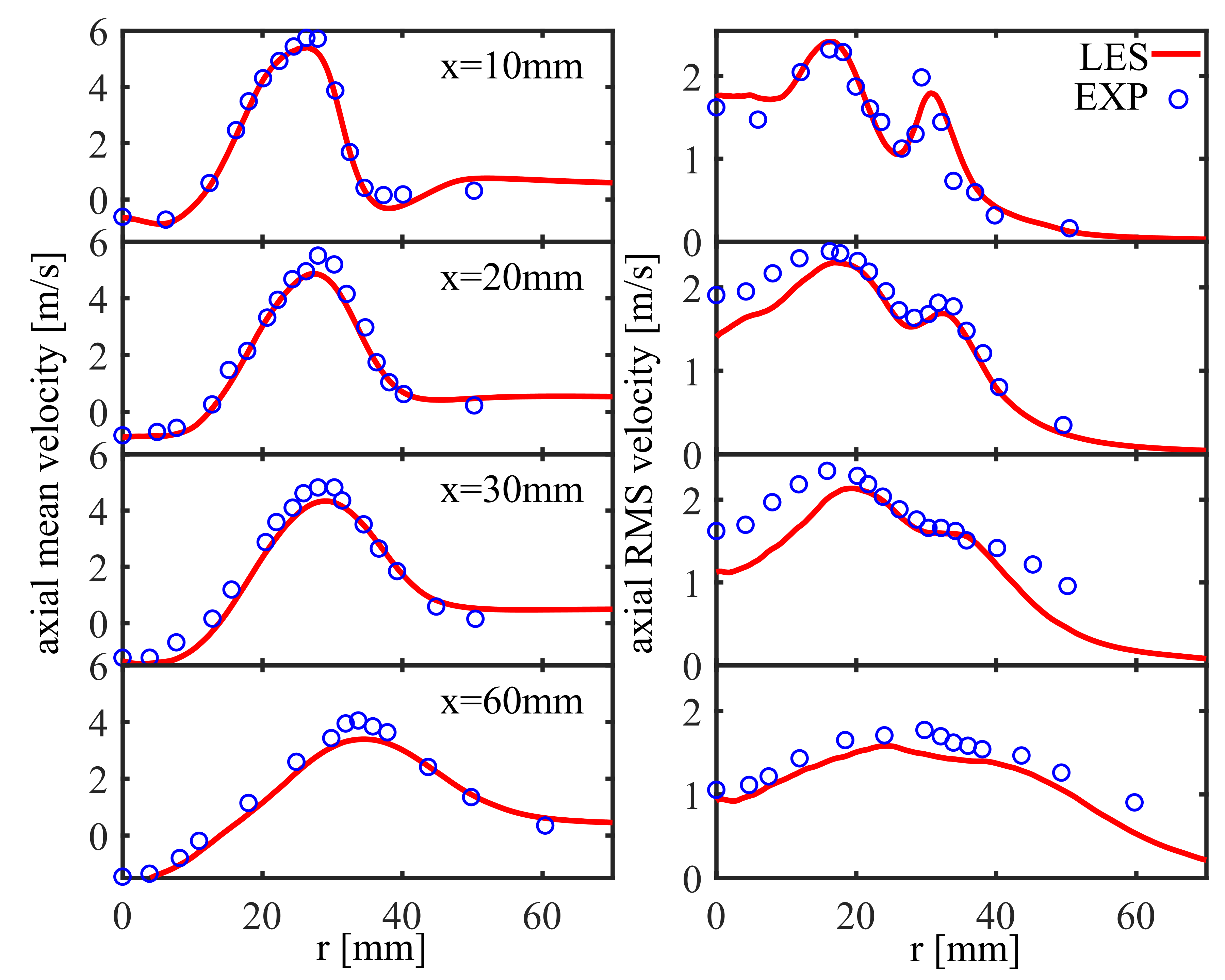}
	\caption{Radial profiles of mean and fluctuating axial velocities at various axial positions for the isothermal case compared to experimental results.}
    \label{fig:axial_cold}
\end{figure}

\begin{figure}[htb]
	\centering
	\includegraphics[width=0.8\textwidth]{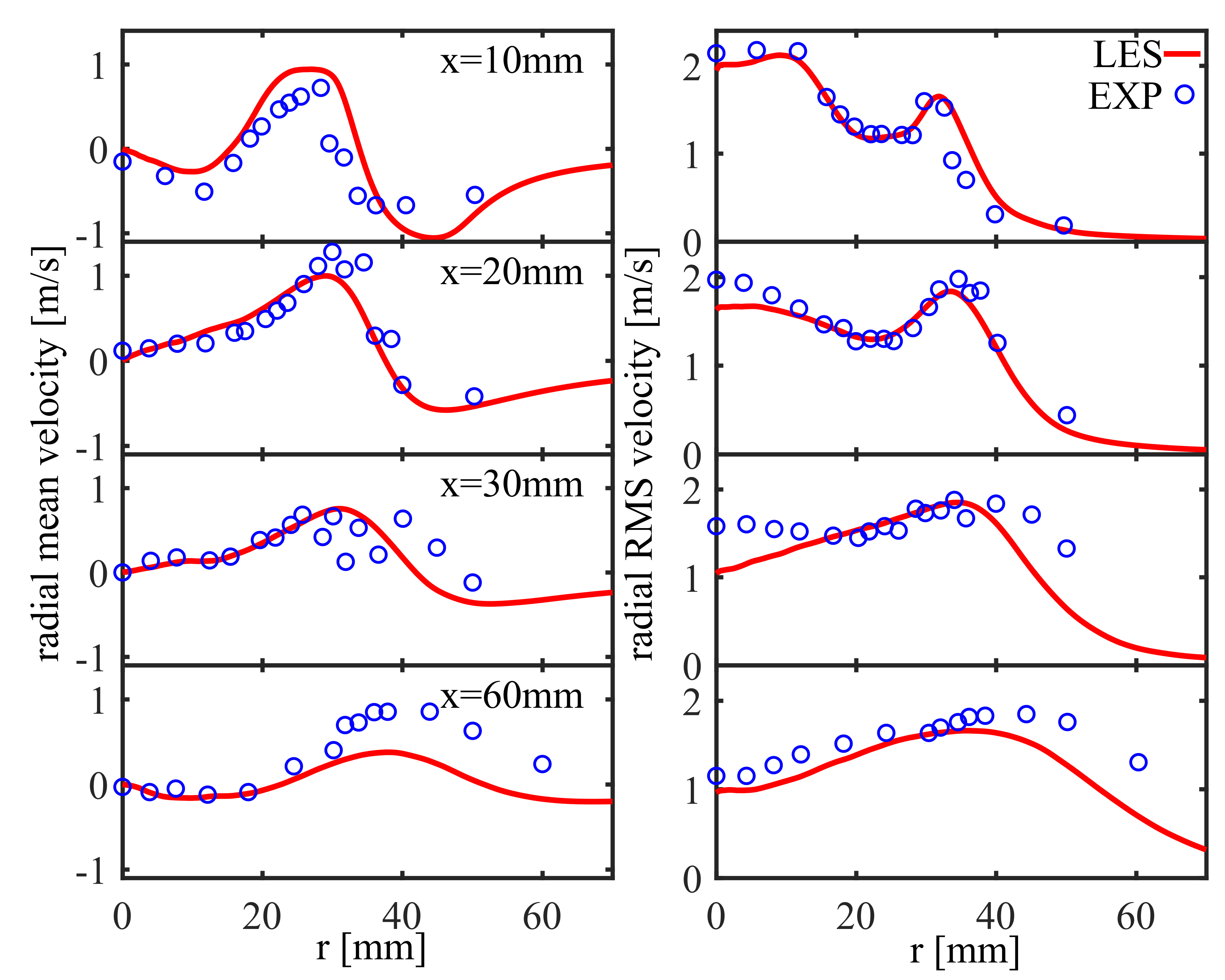}
	\caption{Radial profiles of mean and fluctuating radial velocities at various axial positions for the isothermal case compared to experimental results.}
    \label{fig:radial_cold}
\end{figure}

\begin{figure}[htb]
	\centering
	\includegraphics[width=0.8\textwidth]{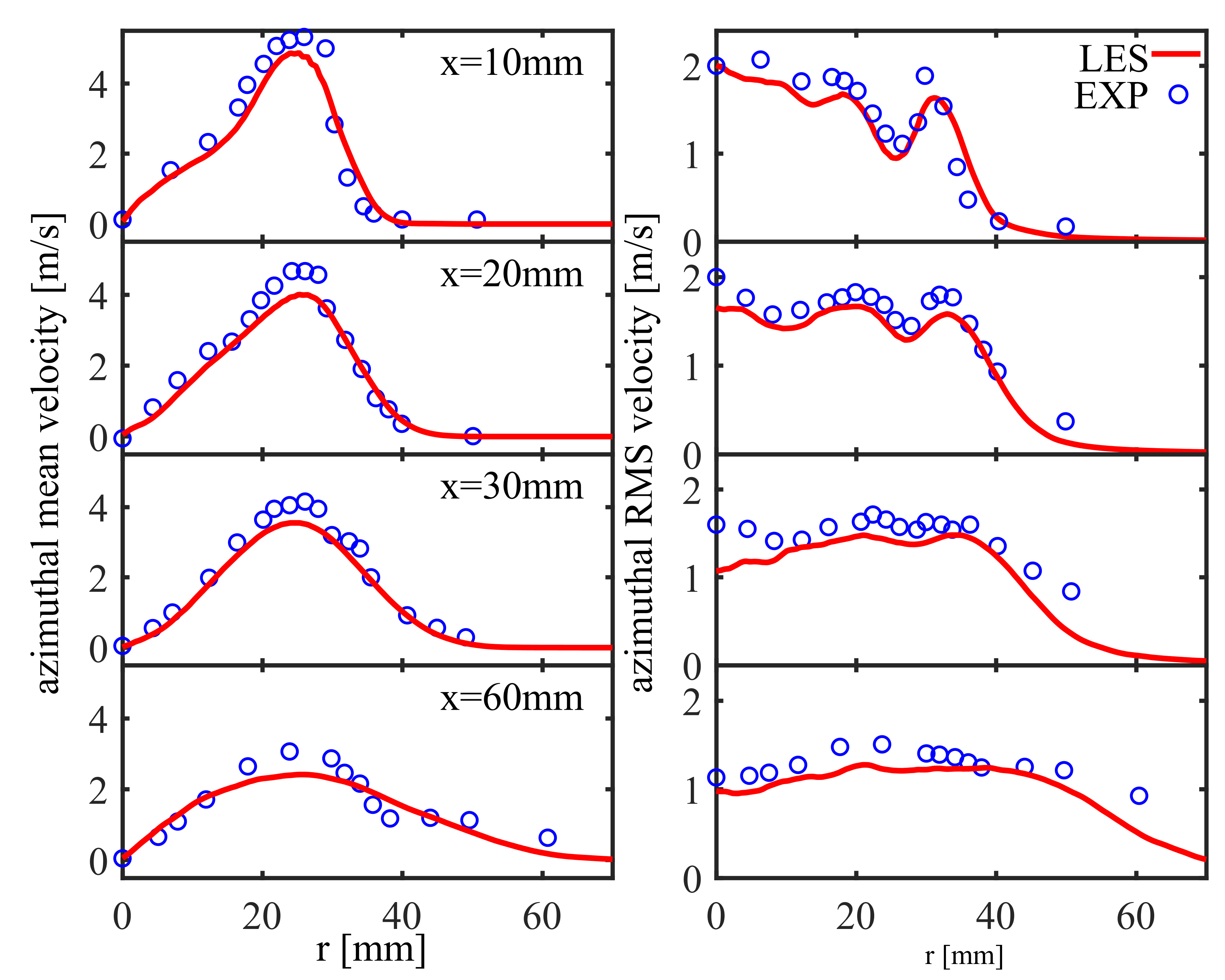}
	\caption{Radial profiles of mean and fluctuating azimuthal velocities at various axial positions for the isothermal case compared to experimental results.}
    \label{fig:azimuthal_cold}
\end{figure}

\subsection{Reactive flow simulation}
\label{sec:Reactive_flow_results}
\begin{figure}[htb]
	\centering
	\includegraphics[width=0.8\textwidth]{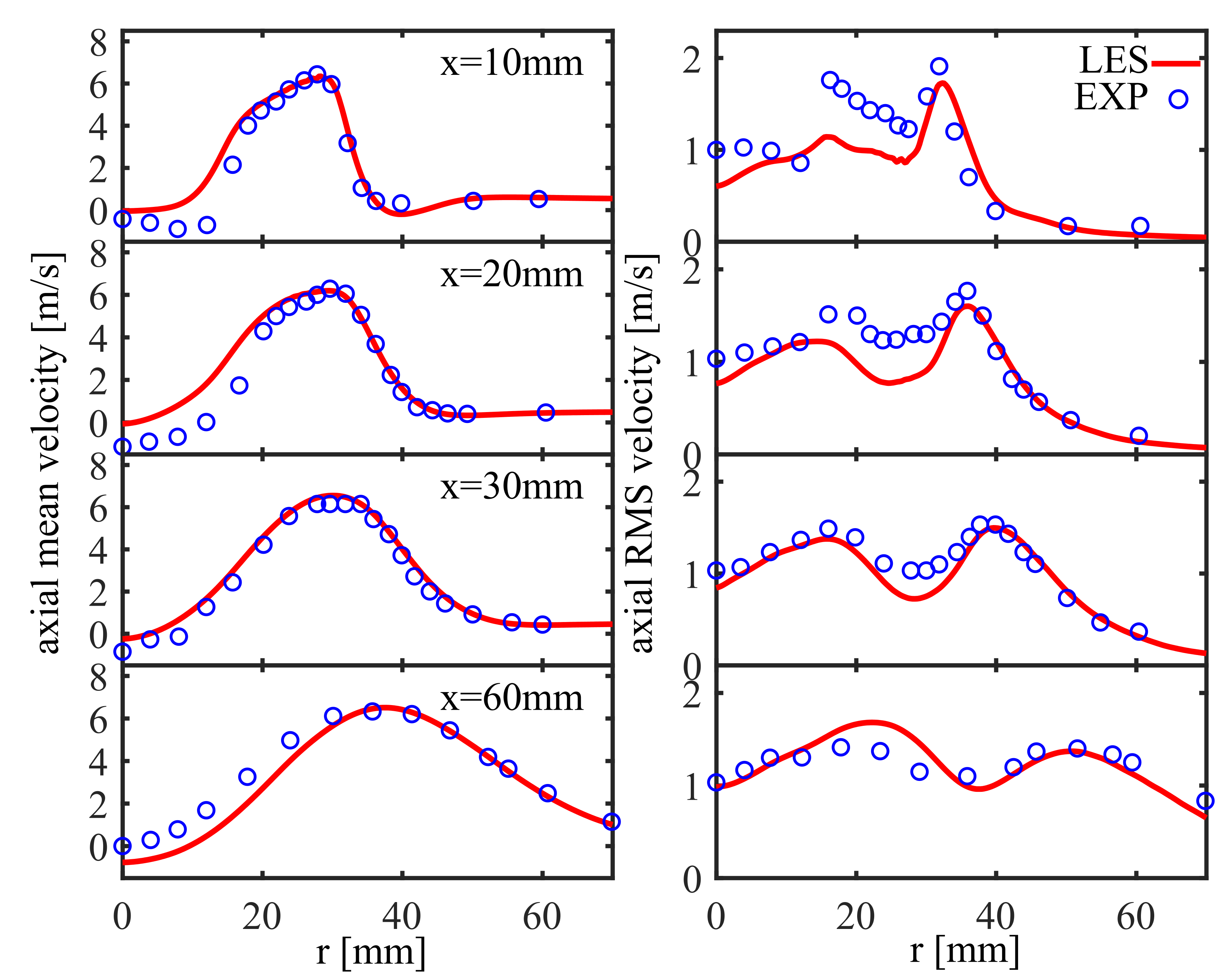}
	\caption{Radial profiles of mean and fluctuating axial velocities at various axial positions for the reacting case compared to experimental results.}
    \label{fig:axial_reac}
\end{figure}

\begin{figure}[htb]
	\centering
	\includegraphics[width=0.8\textwidth]{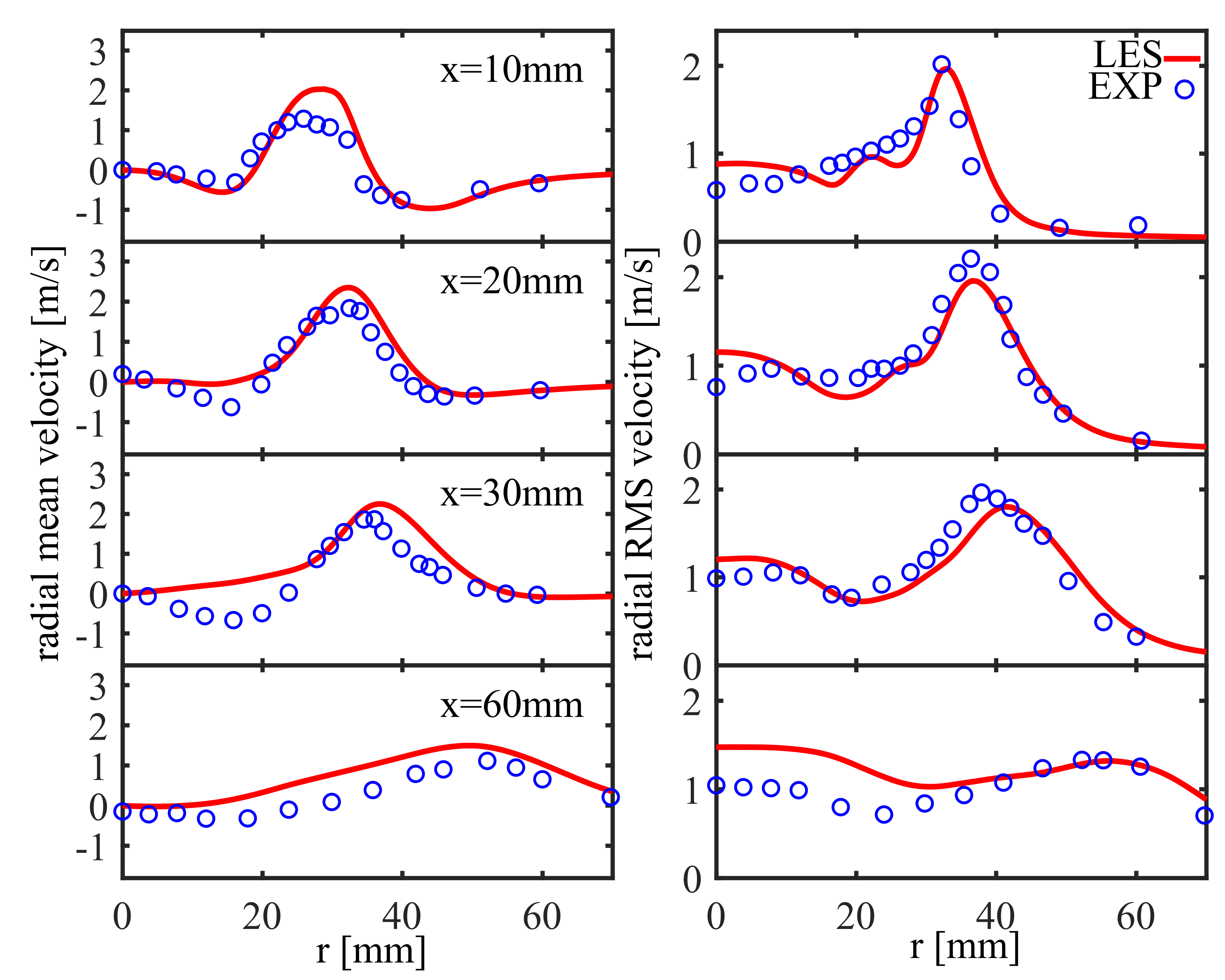}
	\caption{Radial profiles of mean and fluctuating radial velocities at various axial positions for the reacting case compared to experimental results.}
    \label{fig:radial_reac}
\end{figure}

\begin{figure}[htb]
	\centering
	\includegraphics[width=0.8\textwidth]{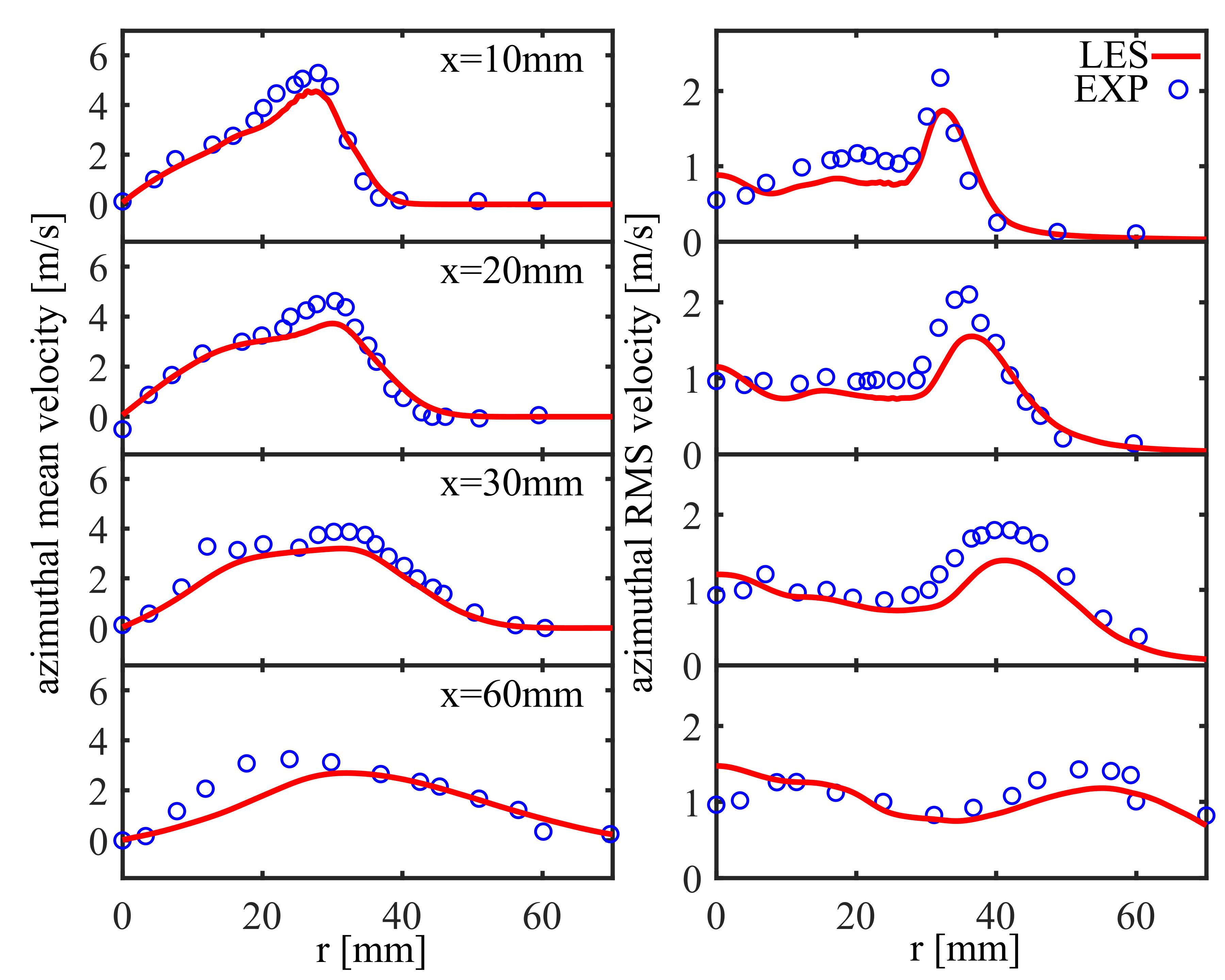}
	\caption{Radial profiles of mean and fluctuating azimuthal velocities at various axial positions for the reacting case compared to experimental results.}
    \label{fig:azimuthal_reac}
\end{figure}

Figures~\ref{fig:axial_reac},~\ref{fig:radial_reac} and~\ref{fig:azimuthal_reac} compare the mean and RMS velocities in the axial, radial, and azimuthal directions, respectively, between the LES predictions and experimental results for the reactive case.
The predicted mean velocities exhibit overall good agreement with the experimental data, and the influence of combustion on the swirling flow field is well reproduced.
Compared to the predictions in the non-reactive case, the mean velocity---especially the radial component---increases within the flame brush due to thermal expansion.
In addition, the size of the recirculation zone, defined by the region of negative axial velocities, is reduced, which is consistent with previous observations~\cite{Hong2015}.
The radial velocity near the centerline of the domain at \(x = 20~\text{mm}\) and \(x = 30~\text{mm}\) is observed to be slightly over-predicted.
This deviation is comparable to findings in previous LES studies~\cite{Jones2012,Butz2015} and might be attributed to uncertainties in the thermal boundary conditions.
Regarding the turbulent statistics, the LES predictions also show satisfactory agreement with the experimental data. 

\begin{figure}[htb]
	\centering
	\includegraphics[width=0.8\textwidth]{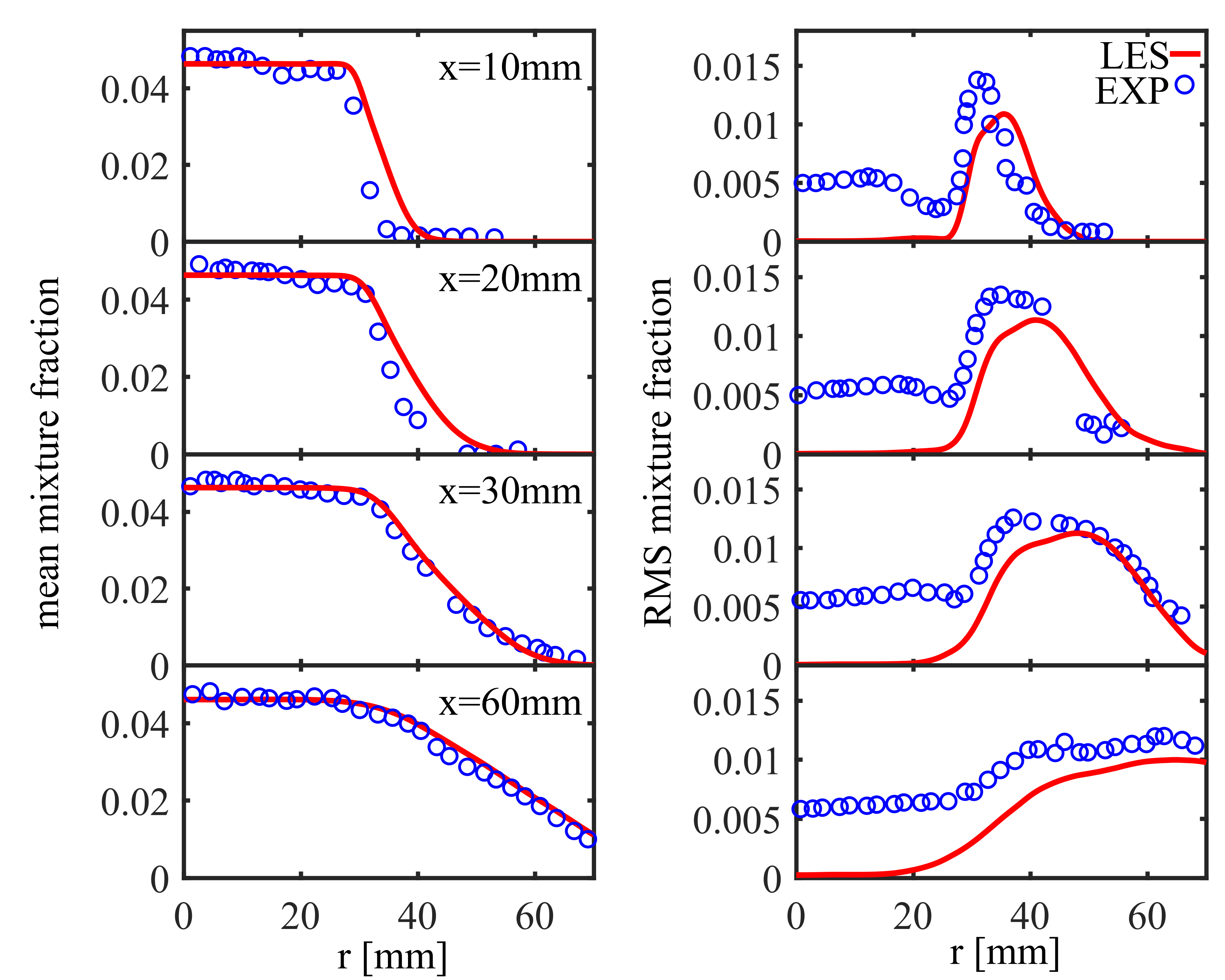}
	\caption{Radial profiles of mean and fluctuating mixture fractions at various axial positions for the reacting case compared to experimental results.}
    \label{fig:mixture_fraction_reac}
\end{figure}

\begin{figure}[htb]
	\centering
	\includegraphics[width=0.8\textwidth]{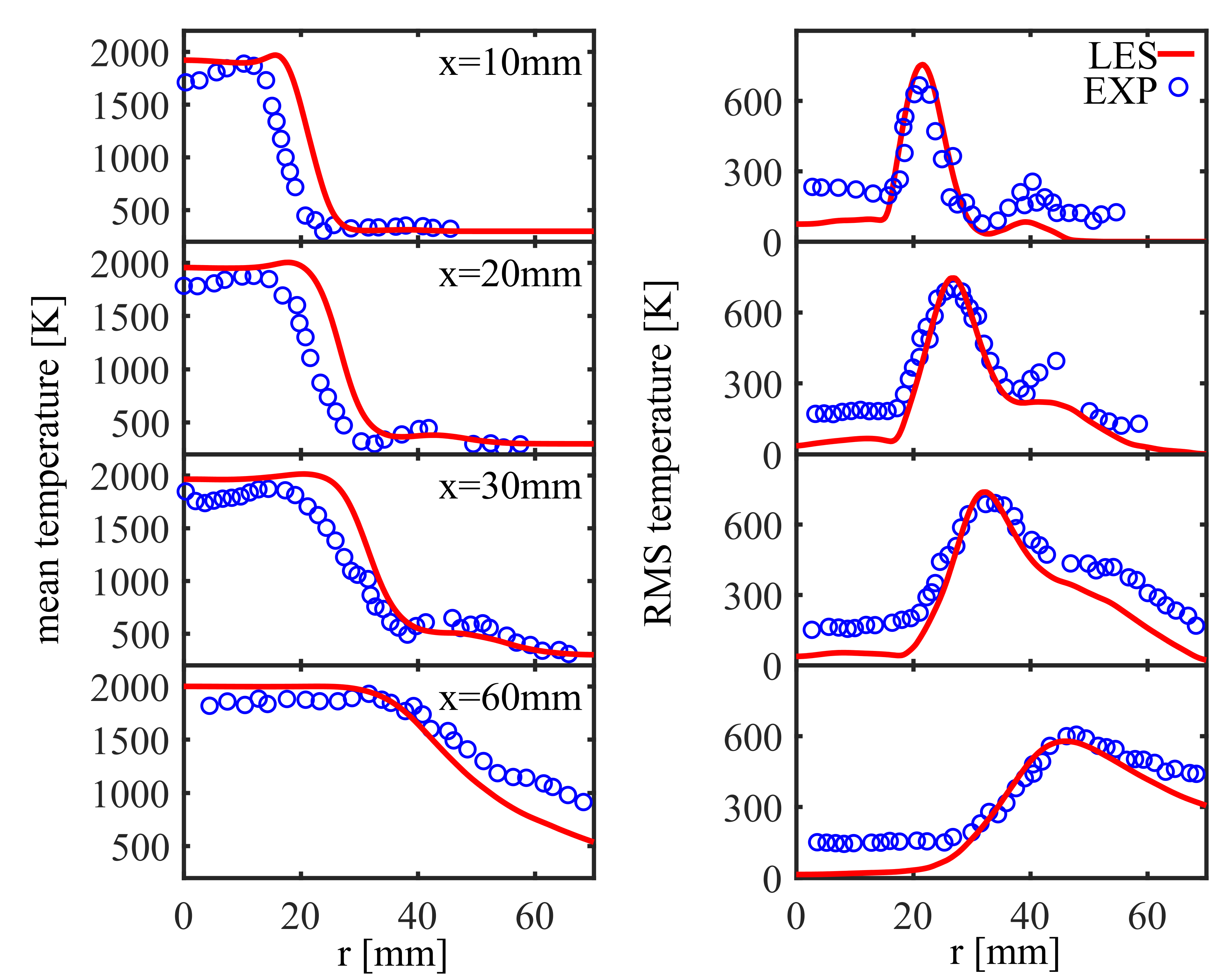}
	\caption{Radial profiles of mean and fluctuating temperatures at various axial positions for the reacting case compared to experimental results.}
    \label{fig:temperature_reac}
\end{figure}

Figures~\ref{fig:mixture_fraction_reac} and~\ref{fig:temperature_reac} present the mean and RMS profiles of the mixture fraction and temperature, respectively, obtained from LES using the FPF model.  
Overall, the LES predictions match well with the experimental measurements.
It is worth noting that the secondary peaks of the mean and fluctuating temperature in the outer shear layer observed experimentally\,(at $x=20$ and $30$~mm for the mean and $x=10-30$~mm for the fluctuations) are successfully captured by the current simulation.  
The underlying mechanism for these secondary peaks and their dependence on the resolved flame thickness are to be discussed in Section~\ref{sec:Flame_thickness_effects}.

\subsection{Effects of resolved flame thickness}
\label{sec:Flame_thickness_effects}

In LES, the resolved flame thickness is determined by the balance of diffusion and sub-filter scale chemical reactions. 
Molecular diffusion naturally tends to thicken the flame front~\cite{Kim2020,Kim2015}. 
Additionally, sub-filter turbulent diffusion, which is commonly modeled via the gradient assumption despite recognized limitations~\cite{Su2022,MacArt2018}, further thickens the resolved flame. 
Therefore, if the reaction-induced thinning effects are not accurately represented in the combustion model, the flame thickness can be over-estimated.

In the FPF method, the steepening effects of the chemical reactions are captured by the model exponent $\gamma$ in Eq.~\eqref{eq:Psi}, which depends on the ratio of the filter size to the laminar flame thickness.
To investigate the impact of the resolved flame thickness on the prediction of flame dynamics in LES, simulation results obtained with $\gamma=2.5$, which is determined from Eq.~\eqref{eq:gamma}, and those with $\gamma=1.0$, where the steepening effects of chemical reactions have been neglected, are compared and analyzed in this sub-section.

\subsubsection{Evolution of resolved flame thickness and wrinklings}
\label{sec:Flame_wrinklings_evolution}

\begin{figure*}[htb]
	\centering
	\includegraphics[width=0.8\textwidth]{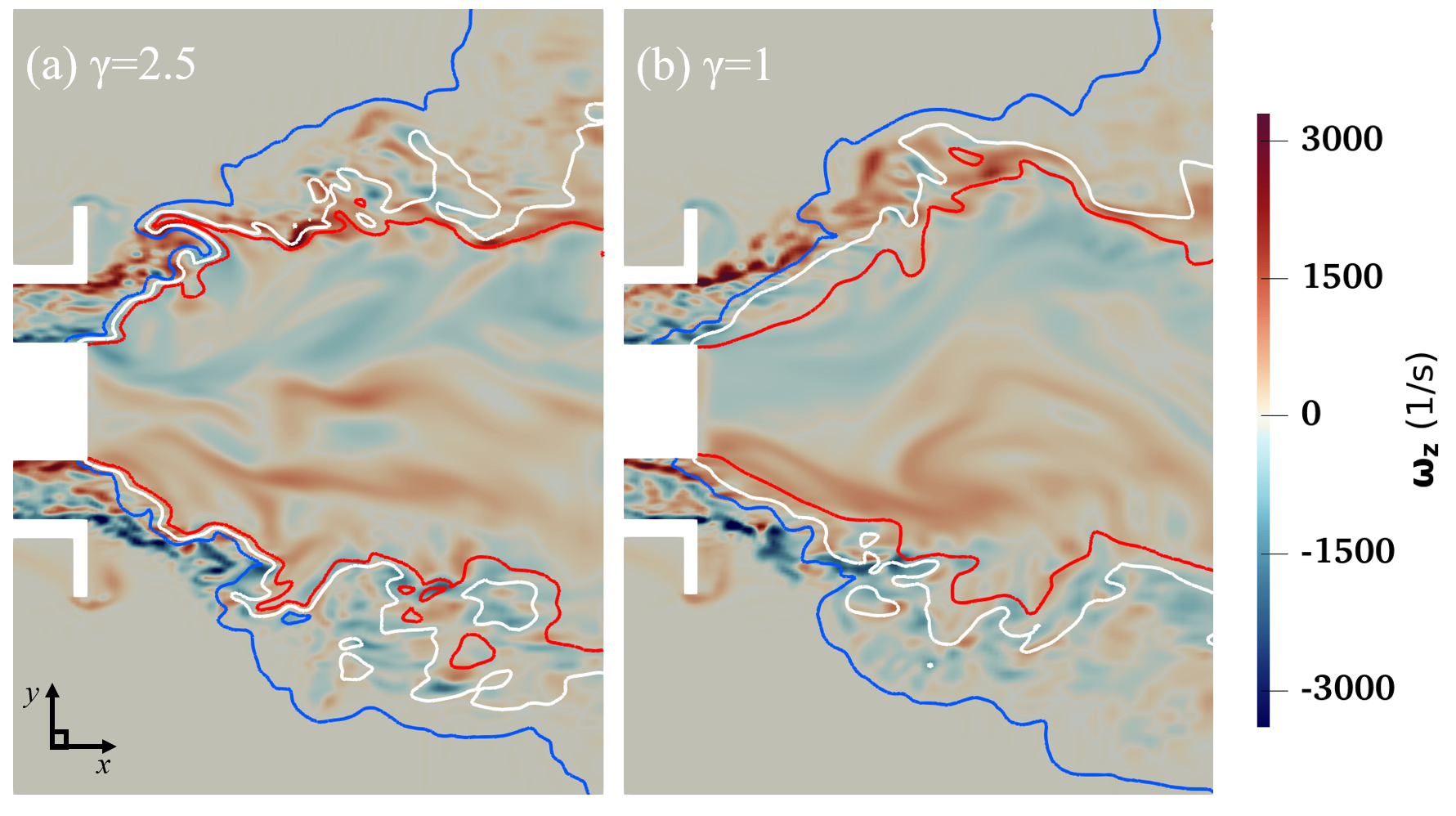}
    \caption{Instantaneous flame structures and contours of the $z$-component of vorticity on the $x$--$y$ cross section obtained with $\gamma=2.5$\,(a) and $\gamma=1.0$\,(b). The blue, white, and red lines represent the iso-lines of $\tilde{c}=0.05$, $0.5$, and $0.95$, respectively.}
    \label{fig:flame_vorticity}
\end{figure*}

Figure~\ref{fig:flame_vorticity} compares the instantaneous flame structures and the cross-sectional normal vorticity fields in the simulations with $\gamma=2.5$ and with $\gamma=1.0$.
The resolved flame structures are represented by the iso-lines of $\tilde{c}=0.05$\,(blue), $\tilde{c}=0.5$\,(white), and $\tilde{c}=0.95$\,(red).
In both cases, the resolved flame becomes increasingly wrinkled downstream.
Distinct differences are evident in the resolved flame structure between the two cases.
Near the nozzle exit, the simulation with $\gamma=2.5$ exhibits a noticeably thinner flame compared to that with $\gamma=1.0$.
Furthermore, this reduced flame thickness with $\gamma=2.5$ is accompanied by significantly more pronounced flame wrinkling.
In terms of the vorticity field, the typical structures of the inner and outer shear layers, characterized by intermittent, intense, and counter-rotating vortices, can be found in Fig.~\ref{fig:flame_vorticity}.
The inner shear layer appears smoother due to the dampening of fine-scale structures by enhanced viscosity induced by combustion heat release.
Additionally, the simulation with $\gamma = 2.5$ exhibits more fragmented vortical structures within the central recirculation zone due to interaction with the more wrinkled flames.

\begin{figure*}[htb]
    \centering
    \includegraphics[width=0.8\textwidth]{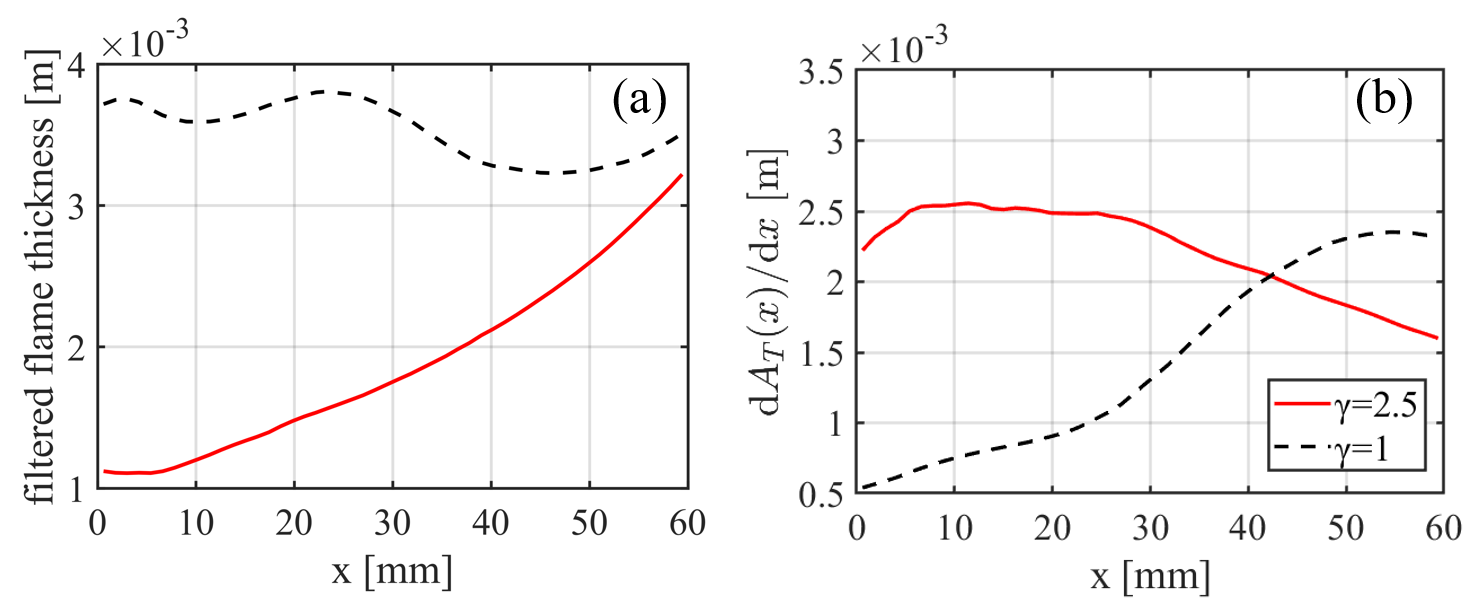}
    \caption{Variation of filtered flame thickness\,(a) and flame surface area\,(b) with axial position for different \(\gamma\) values.}
    \label{fig:filter_average}
\end{figure*}

To quantitatively compare the flame wrinkling obtained with different resolved flame thicknesses, the axial evolution of the averaged filtered flame thickness, $l_{F,\Delta}$, and the growth rate of the flame surface area, $dA_T(x)/dx$, for $\gamma=2.5$ and $\gamma=1.0$ are presented in Fig.~\ref{fig:filter_average}.
$l_{F,\Delta}$ is defined as the reciprocal of the surface-averaged gradient of the Favre-filtered progress variable, $\overline{|\nabla \tilde{c}|}$.
The term $\overline{|\nabla \tilde{c}|}$ is evaluated as:
\begin{equation}
\overline{|\nabla \tilde{c}|} = \frac{\int_S \int_0^T|\nabla \tilde{c}| |\nabla \tilde{c}| \delta(\tilde{c}-c^{\ast}) dsdt}{\int_S \int_0^T |\nabla \tilde{c}| \delta(\tilde{c}-c^{\ast})\, dsdt},
\label{eq:mean_gradient}
\end{equation}
where $S$ denotes the cross-section of the computational domain normal to the axial coordinate $\text{x}$, and $T$ is the time period used for evaluating statistics.
$\delta(\tilde{c}-c^{\ast})$ is the Dirac delta function identifying the flame surface at $\tilde{c}=c^{\ast}$~\cite{Veynante2005}.
Here, $c^{\ast}$ is set to $0.5$, following previous studies~\cite{Su2025,Su2025fractal}.
The cumulative flame surface area, $A_T(\text{x})$, is calculated as,
\begin{equation}
A_T(\text{x})=\frac{1}{T} \int_0^T \int_0^{\text{x}} \int_S|\nabla \tilde{c}| \delta(\tilde{c}-c^{\ast}) ds dx dt.
\end{equation}

Figure~\ref{fig:filter_average}\,(a) corroborates the qualitative observations of the resolved flame thickness from Fig.~\ref{fig:flame_vorticity}.
For $x<20$~mm, the value of $l_{F,\Delta}$ obtained in the case with $\gamma = 1.0$ is approximately three times that of the $\gamma = 2.5$ case.
In the simulation with $\gamma=2.5$, $l_{F,\Delta}$ increases gradually with axial distance, particularly for $x>20$~mm.
This thickening arises from enhanced mixing processes, which reduce the intensity of local chemical reactions.
In contrast, $l_{F,\Delta}$ remains relatively constant in the simulation with $\gamma=1.0$.
Due to the absence of the steepening mechanisms, the modeled diffusion processes seem to dominate over the fuel stratification in thickening the filtered flame in LES.

Figure~\ref{fig:filter_average}\,(b) further confirms that increased flame thickness leads to a reduction in the resolved flame surface area.
At $x=30$~mm, $dA_T(x)/dx$ for $\gamma=1.0$ is approximately half that of the case with $\gamma=2.5$.
In LES of turbulent premixed combustion where the flamelet assumption is valid, the thickness of the resolved flame poses a lower limit on the smallest scale of the resolved flame wrinklings.
Consequently, if the filter size—rather than the actual resolved flame thickness—is used to model sub-filter wrinkling, the total consumption rate may be significantly underestimated, especially under conditions where chemical steepening effects are not adequately captured.

\subsubsection{Secondary temperature peaks and flame pockets}
\label{sec:Secondary_temperature_peak}

\begin{figure}[!ht]
	\centering
	\includegraphics[width=0.8\textwidth]{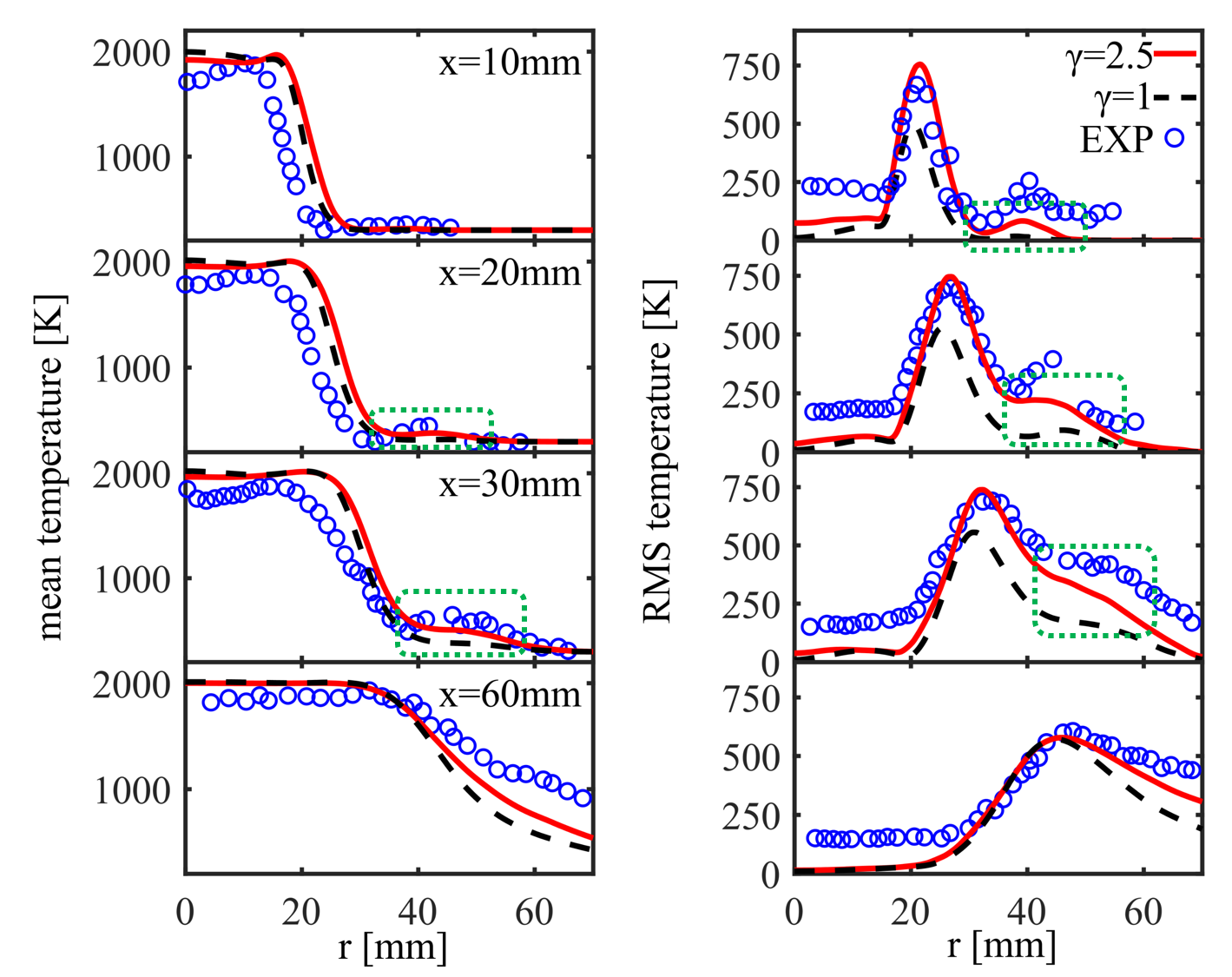}
\caption{Comparison of radial profiles of mean and fluctuating temperatures at different axial positions between the simulations with $\gamma=2.5$ and $\gamma=1.0$. The green dotted boxes highlight the secondary peaks or plateaus in the profiles.}
    \label{fig:temperature_gamma}
\end{figure}

Figure~\ref{fig:temperature_gamma} compares the mean and fluctuating temperature profiles obtained from the simulations with $\gamma=2.5$ and $\gamma=1.0$.
The width of the high-temperature zone is noticeably narrower in the simulation with $\gamma=1.0$ compared to that with $\gamma=2.5$.
This discrepancy is attributed to the underestimation of the total consumption rate when the resolved flame thickness is over-estimated with $\gamma=1.0$, as discussed in Section~\ref{sec:Flame_wrinklings_evolution}.
Regarding the fluctuating temperature profiles, the major peak near the inner shear layer is considerably better reproduced in the simulation with $\gamma=2.5$.
The temperature fluctuations near the inner shear layer mainly arise from the intermittent behavior of the turbulent flames, which leads to the alternating presence of unburned and burned states at a fixed location.
With the model parameter $\gamma$ determined from the physics-based models, the thickness of the resolved flame is well predicted, thereby enabling the accurate capture of the flapping motions of instantaneous flame fronts.
In contrast, the over-estimation of the resolved flame thickness in the simulation with $\gamma=1.0$ results in an under-prediction of this intermittent behavior.
This highlights the capability of the FPF method in predicting the small-scale flame dynamics, consistent with findings reported in previous LES of Bunsen flames~\cite{Su2025}.

\begin{figure*}[!ht]
	\centering
	\includegraphics[width=0.8\textwidth]{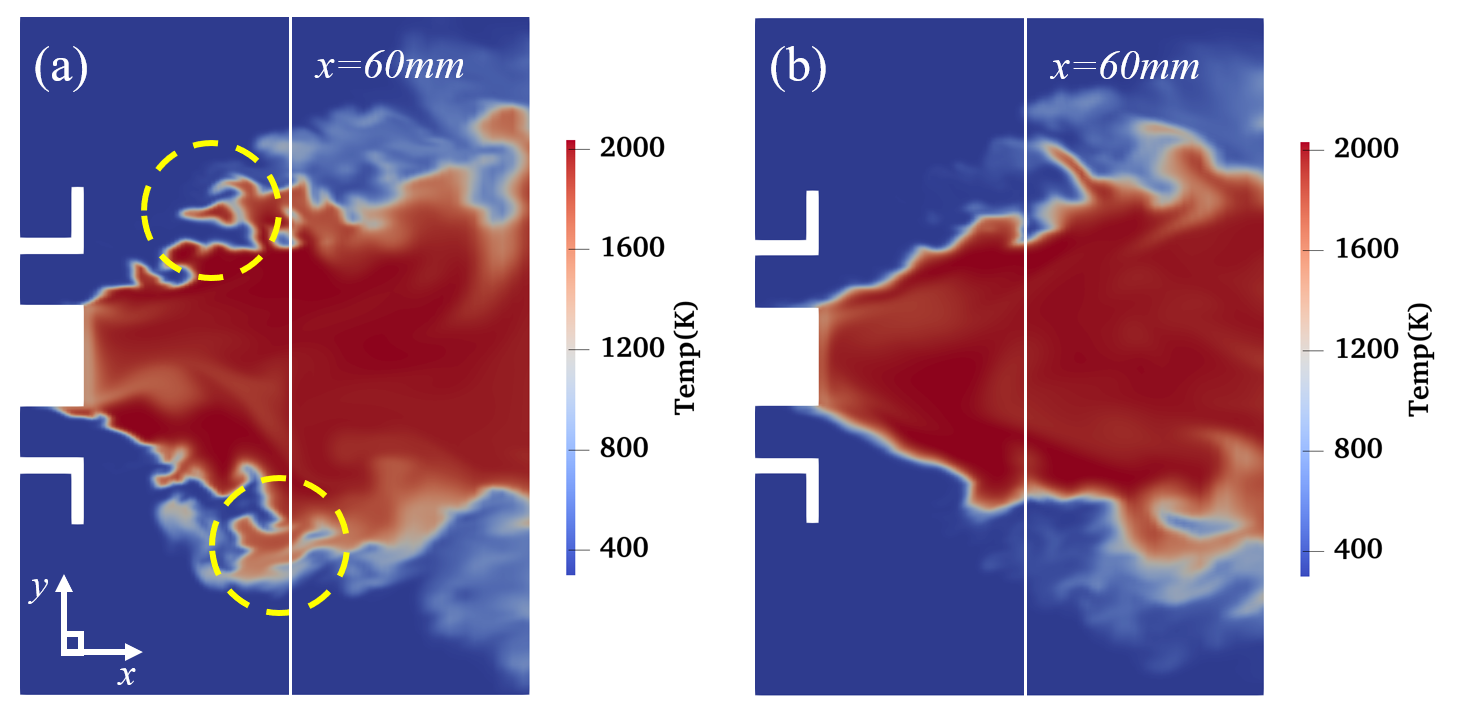}
	\caption{Representative temperature contours obtained with $\gamma=2.5$\,(a) and $\gamma=1.0$\,(b). The yellow dashed circles mark the isolated flame pockets and peninsulas.}
    \label{fig:temperature_contour_gamma}
\end{figure*}

In addition to the major peak, the secondary peak of the fluctuating temperature near the outer shear layer is also noticeably better reproduced when the thickness of the resolved flame is well maintained using $\gamma=2.5$, as highlighted by the green dotted boxes in Fig.~\ref{fig:temperature_gamma}.
To elucidate the origin of these fluctuations, Fig.~\ref{fig:temperature_contour_gamma} presents the instantaneous temperature contours.
Figure~\ref{fig:temperature_contour_gamma}\,(a) shows that this peak is attributed to the presence of flame structures---specifically isolated pockets~\cite{Tyagi2020} or connected peninsulas---that detach or extend from the main flame surface and penetrate into the outer shear layer.
In contrast, such structures are absent in the simulation with $\gamma=1$ near the outer shear layer in the region $x<60$\,mm.
This is observed neither in the specific instance in Fig.~\ref{fig:temperature_contour_gamma}\,(b) nor at any other time instant.
 
\begin{figure*}[!ht]
	\centering
	\includegraphics[width=0.8\textwidth]{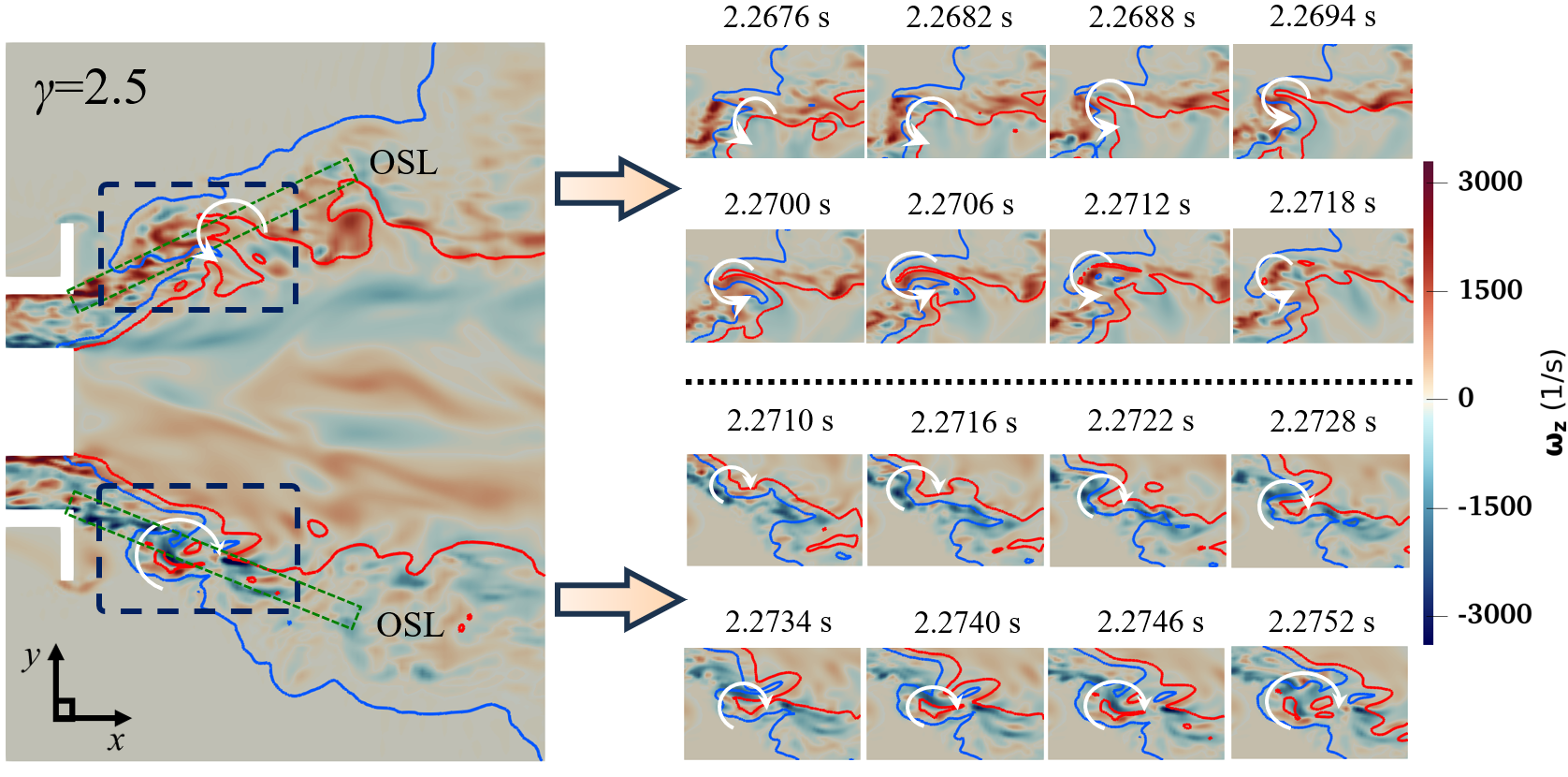}
	\caption{Temporal evolution of resolved eddy-flame interactions within the region $5\,\text{mm}<x<60\,\text{mm}$ and $25\,\text{mm}<|y|<60\,\text{mm}$ on the $x$--$y$ plane for the case with $\gamma=2.5$. Blue and red solid lines denote the iso-lines of $\tilde{c}=0.05$ and $0.95$, respectively. The background color represents the $z$-component of vorticity, $\omega_z$. Black dashed boxes on the left indicate the regions corresponding to the sequential interactions shown on the right. Green dashed boxes denote the approximate position of the outer shear layer\,(OSL).}
    \label{fig:evolution_gamma_2.5}
\end{figure*}

\begin{figure*}[!ht]
	\centering
	\includegraphics[width=0.8\textwidth]{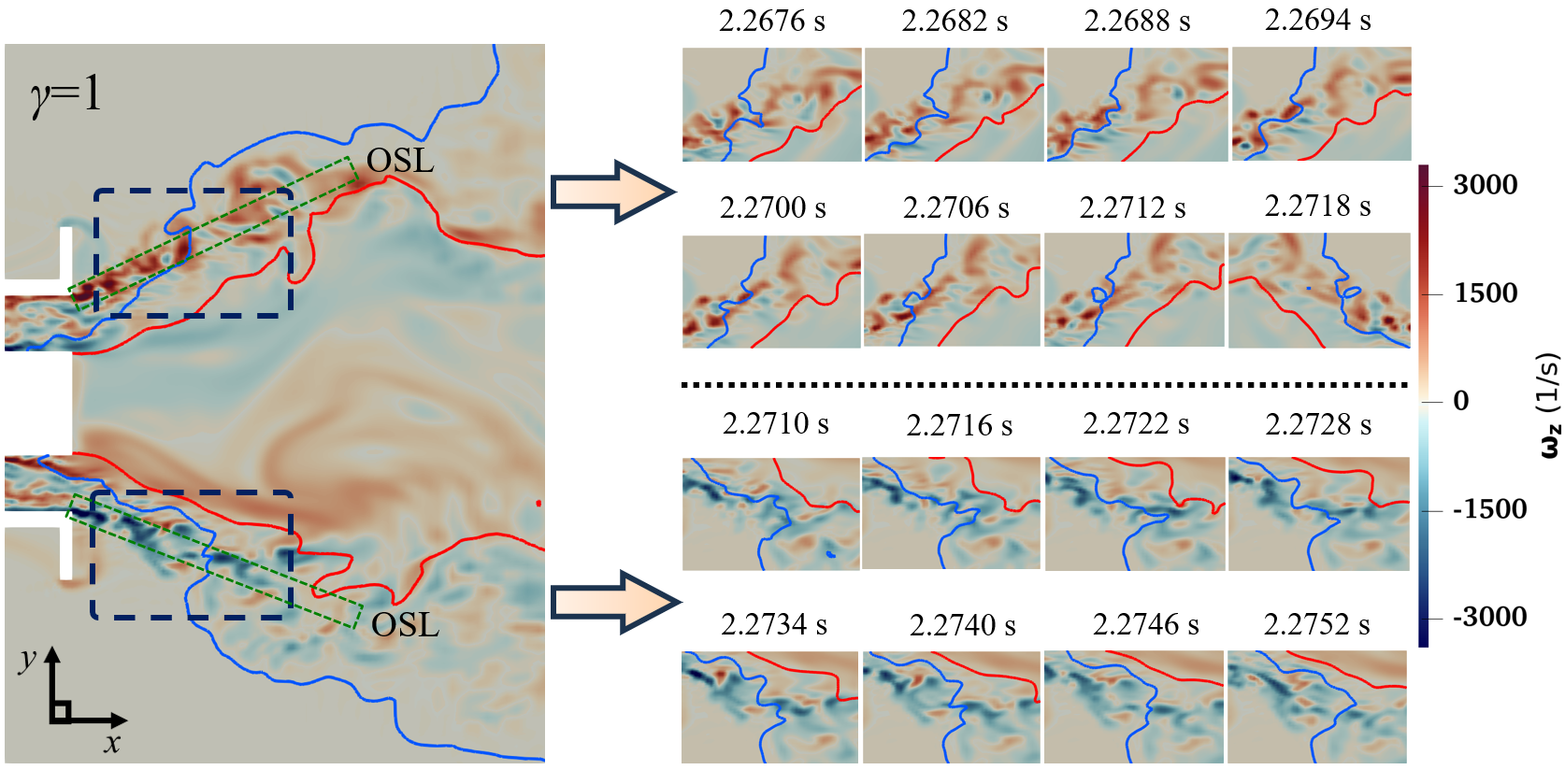}
	\caption{Temporal evolution of resolved eddy-flame interactions within the region $5\,\text{mm}<x<60\,\text{mm}$ and $25\,\text{mm}<|y|<60\,\text{mm}$ on the $x$--$y$ plane for the case with $\gamma=1.0$. The line styles, background coloring, and box definitions are the same as in Fig.~\ref{fig:evolution_gamma_2.5}.}
    \label{fig:evolution_gamma_1}
\end{figure*}

To investigate the formation of the flame pockets and, more importantly, the effects of the resolved flame thickness on their prediction in LES of swirling flames, the spatial-temporal interactions between the resolved flame and the vortical structures are presented in Fig.~\ref{fig:evolution_gamma_2.5} and Fig.~\ref{fig:evolution_gamma_1} for $\gamma=2.5$ and $\gamma=1.0$, respectively.
The red and blue lines denote the iso-lines of \(\tilde{c} = 0.95\) and \(\tilde{c} = 0.05\), respectively.
The time interval between frames in both figures is set to $0.0006$~s.
As observed in Fig.~\ref{fig:evolution_gamma_2.5}, the vortical structures in the outer shear layer evolves in the streamwise direction.
These structures rotate in the counter-clock direction on the upper panel and in the clockwise direction in the lower panel.
Simultaneously, the flame front tilts outward in the radial direction driven by flame propagation and the centrifugal force.
Near the location where the tilted flame intersects with the outer shear layer, relatively large-scale rotating vortices penetrate the filtered flame and stretch the trailing edge of the flame front.
This induces flame roll-up and leads to the formation of peninsulas and isolated pockets of hot gases.
Due to the rotation of the vortices, the isolated pockets can be transported towards the co-flow side of the outer shear layer and become trapped in the low-velocity region.
This phenomenon is responsible for the secondary peak of the temperature observed in Fig.~\ref{fig:temperature_gamma}.
The trapped flame pockets in the TECFLAM burner was attributed to the radial flapping motion of the main reactant stream by Butz et al.~\cite{Butz2015}.
However, the present temporal sequences of the flame-vortex interactions support a new view that relatively large-scale rotating eddies in the outer shear layer stretch the flame, causing the formation and subsequent trapping of these isolated pockets.

The prediction of the interaction between the resolved flame and turbulent eddies near the outer shear layer strongly depends on the resolved flame thickness.
As shown in Fig.~\ref{fig:evolution_gamma_1}, with the steepening effects neglected for $\gamma=1.0$, there is no intersection between the outer shear layer and the resolved flame.
At locations of $x>30$~mm, although rotating eddies penetrate the resolved flame, only the leading edge is stretched.
The trailing edge of the resolved flame remains unperturbed, as it is detached from the outer shear layer.
Consequently, no flame pockets are observed in the simulations with $\gamma=1.0$.
As shown in Fig.~\ref{fig:filter_average}, over-prediction of the resolved flame thickness leads to significant under-prediction of the resolved flame surface area.
When the filter size is used to model the sub-filter consumption rate, this results in an under-estimation of the turbulent flame speed.
This, in turn, leads to an under-predicted flame angle\,(the angle between the center line and the flame brush), causing the resolved flame to detach from the outer shear layer and resulting in the failure to predict flame pocket formation in that region.

\section{Conclusions}
\label{sec:Conclusions}
This study presents LES of the turbulent stratified and non-adiabatic TECFLAM swirl-stabilized flame with the extended FPF method. 
The primary objectives were to extend the FPF approach to complex swirling combustion and to investigate the effects of the resolved flame thickness on the predicted flame dynamics in LES. 
The FPF method is designed to mitigate spurious propagation in under-resolved flames and reproduce reaction characteristics of filtered flame fronts. 
While previous studies demonstrated its performance in canonical combustion problems, this work marks its first application to LES of turbulent swirling combustion involving fuel stratification, non-adiabaticity, and strong vortex-flame interactions. 
In this work, the FPF method is integrated with a non-adiabatic flamelet sub-model and a refined sub-filter flame speed estimation that resolves the inconsistency arising from heat-release effects on local sub-filter turbulence.

\textit{A posteriori} validation demonstrates that the LES predictions achieve satisfactory good agreement with experimental measurements regarding velocity, mixture fraction, and temperature distributions. 
In particular, LES successfully reproduces the peaks of velocity fluctuations and the secondary temperature peaks. 
These results confirm that the extended FPF method is capable of accurately capturing flame dynamics in complex swirling combustors with heat loss effects at relatively low Karlovitz numbers.

Analysis of the simulation results reveals that the relatively large-scale rotating eddies in the outer shear layer stretch the swirling flame, causing the formation and subsequent trapping of isolated flame pockets or peninsula structures, which manifests as the observed secondary temperature peaks.
Notably, the prediction of this phenomenon is shown to be strongly dependent on the resolved flame thickness.
When the filter size is used for modeling the sub-filter wrinkling in LES, an over-predicted flame thickness results in an under-estimation of the overall consumption rate.
This alters the relative position between the flame front and the outer shear layer, which in turn significantly affects the generation of isolated flame structures. 
These findings underscore the importance of the accurately reproducing the resolved flame thickness--or equivalently, properly modeling chemical steepening effects--in LES of turbulent swirl-stabilized flames.

\section*{Credit authorship contribution statement}

\textbf{Ruochen Guo:} Writing – original draft, Visualization, Investigation, Formal analysis, Data curation. \textbf{Yunde Su:} Writing – review \& editing, Supervision, Conceptualization, Resources, Methodology, Funding acquisition, Formal analysis. \textbf{Yuewen Jiang:} Writing – review \& editing, Supervision, Resources, Funding acquisition, Project administration.

\section*{Declaration of Competing Interest}
The authors declare that they have no known competing financial interests or personal relationships that could have appeared to influence the work reported in this paper.

\section*{Acknowledgments}
We gratefully acknowledge S. Zhang for providing the inflow data used in the present simulations.
This work has been supported by the Open Research Program of National Key Laboratory of Fundamental Algorithms and Models for Engineering Simulation\,(No. 00304454A8001), the Fundamental Research Funds for the Central Universities\,(No. 1082204112K92), Sichuan University Interdisciplinary Innovation Fund\,(No. 1082204112L82), and National Natural Science Foundation of China\,(No. 52476041).

\section*{Appendix A. Grid convergence study}
\label{sec:Grid_convergence}
\begin{figure}[!ht]
	\centering
	\includegraphics[width=0.8\textwidth]{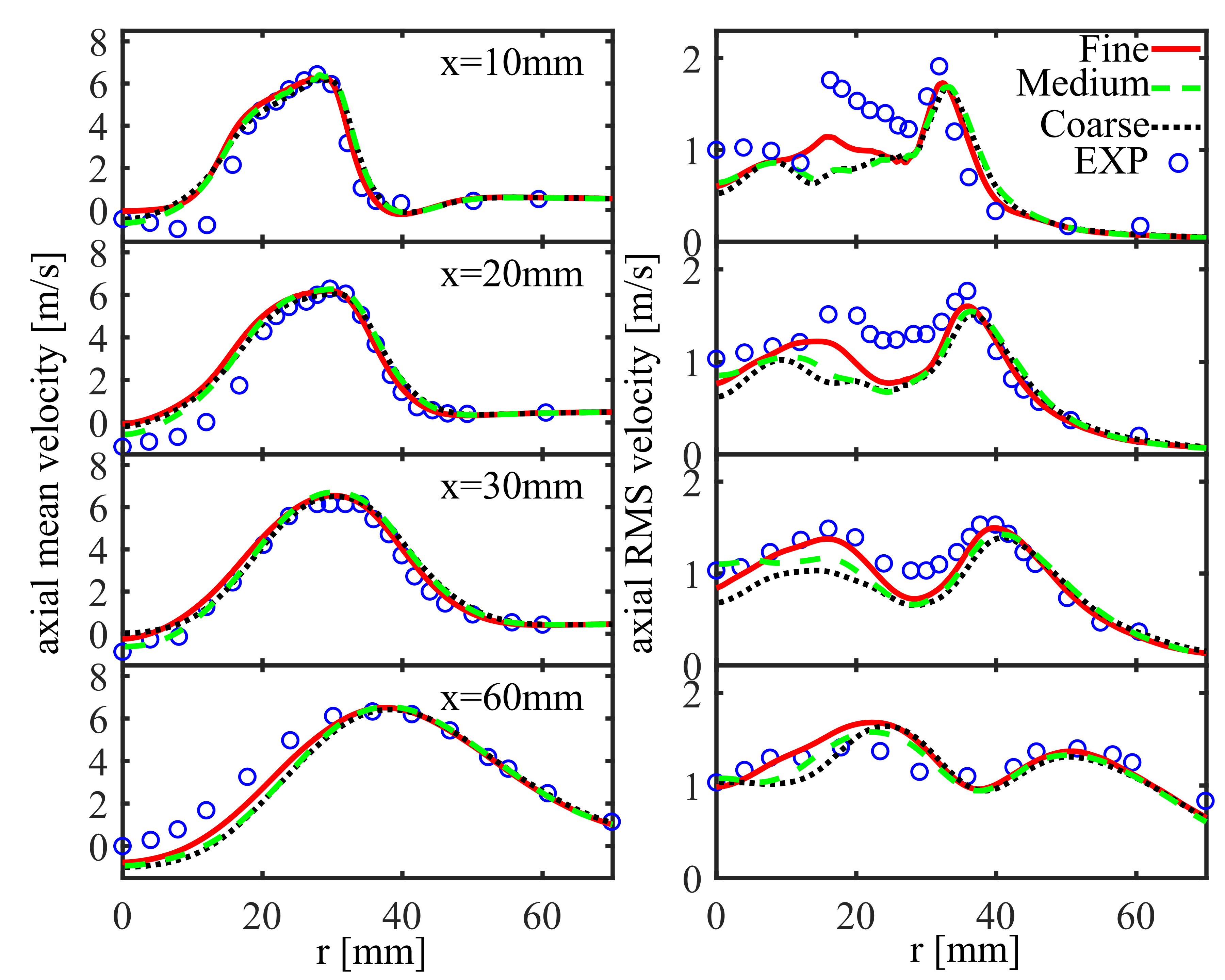}
	\caption{Comparison of radial profiles of mean and fluctuating axial velocities at different axial positions for the coarse, medium and fine grids.}
    \label{fig:convergence_reac_axial}
\end{figure}

\begin{figure}[!ht]
	\centering
	\includegraphics[width=0.8\textwidth]{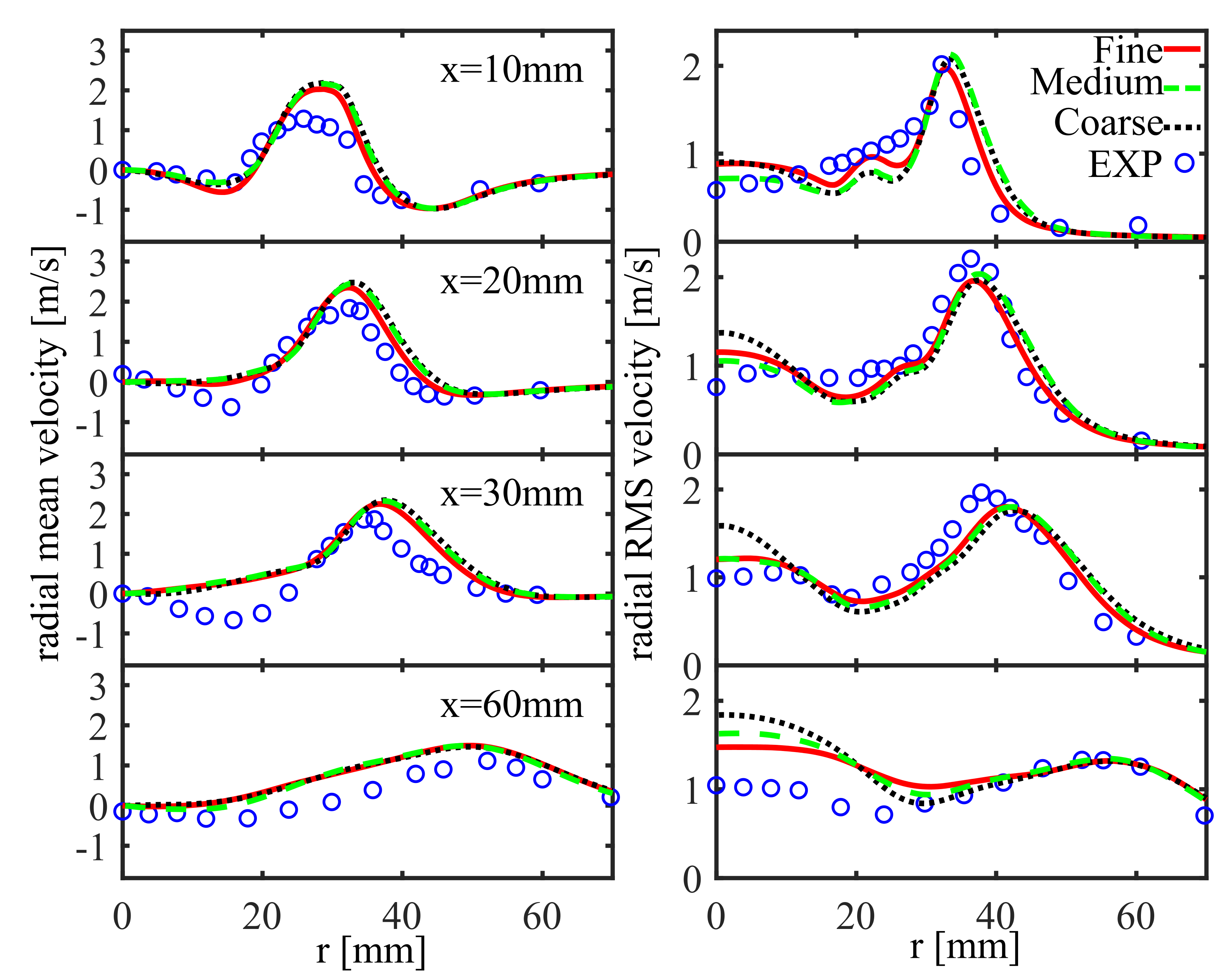}
	\caption{Comparison of radial profiles of mean and fluctuating radial velocities at different axial positions for the coarse, medium and fine grids.}
    \label{fig:convergence_reac_radial}
\end{figure}

\begin{figure}[!ht]
	\centering
	\includegraphics[width=0.8\textwidth]{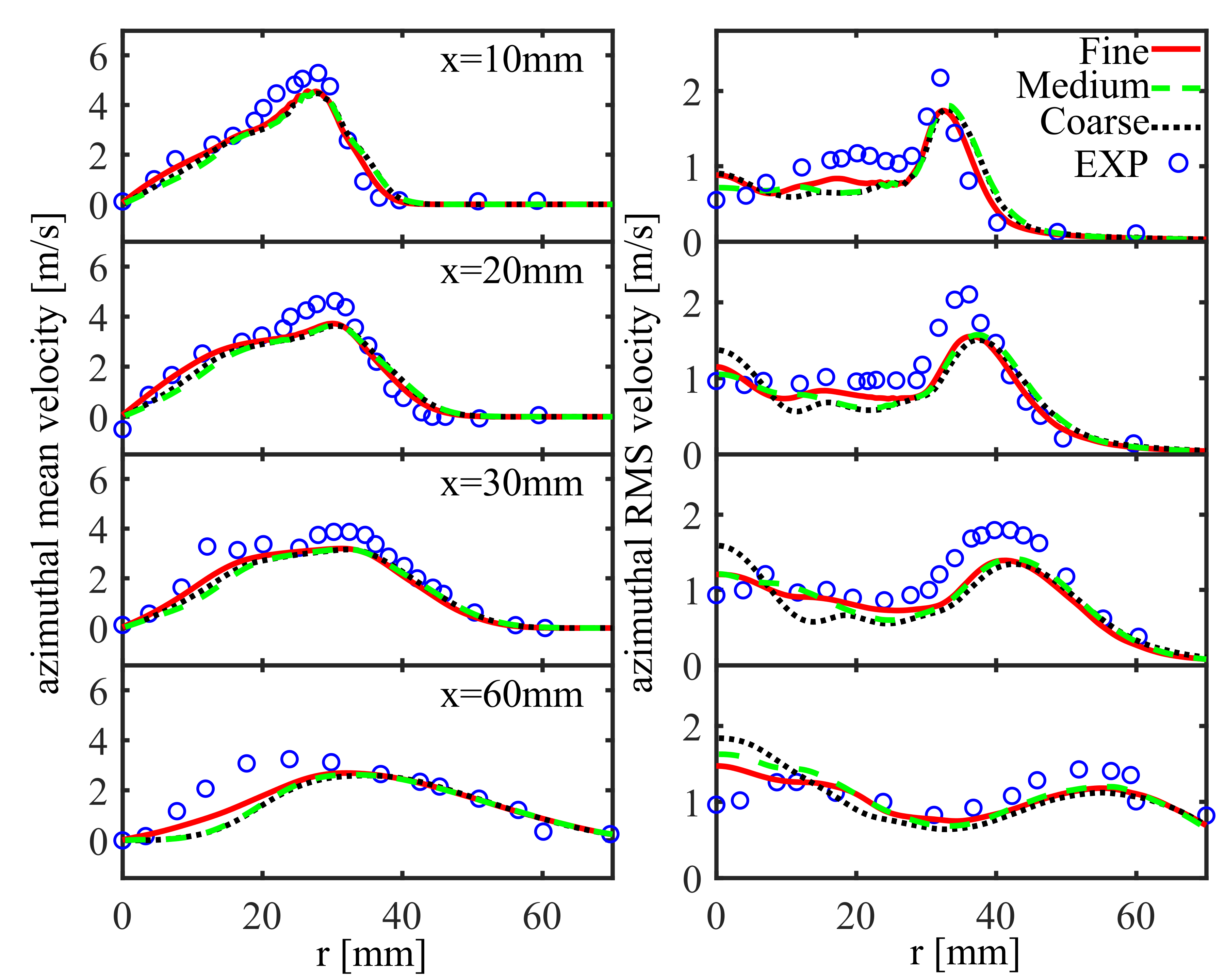}
	\caption{Comparison of radial profiles of mean and fluctuating azimuthal velocities at different axial positions for the coarse, medium and fine grids.}
    \label{fig:convergence_reac_azimuthal}
\end{figure}

\begin{figure}[!ht]
	\centering
	\includegraphics[width=0.8\textwidth]{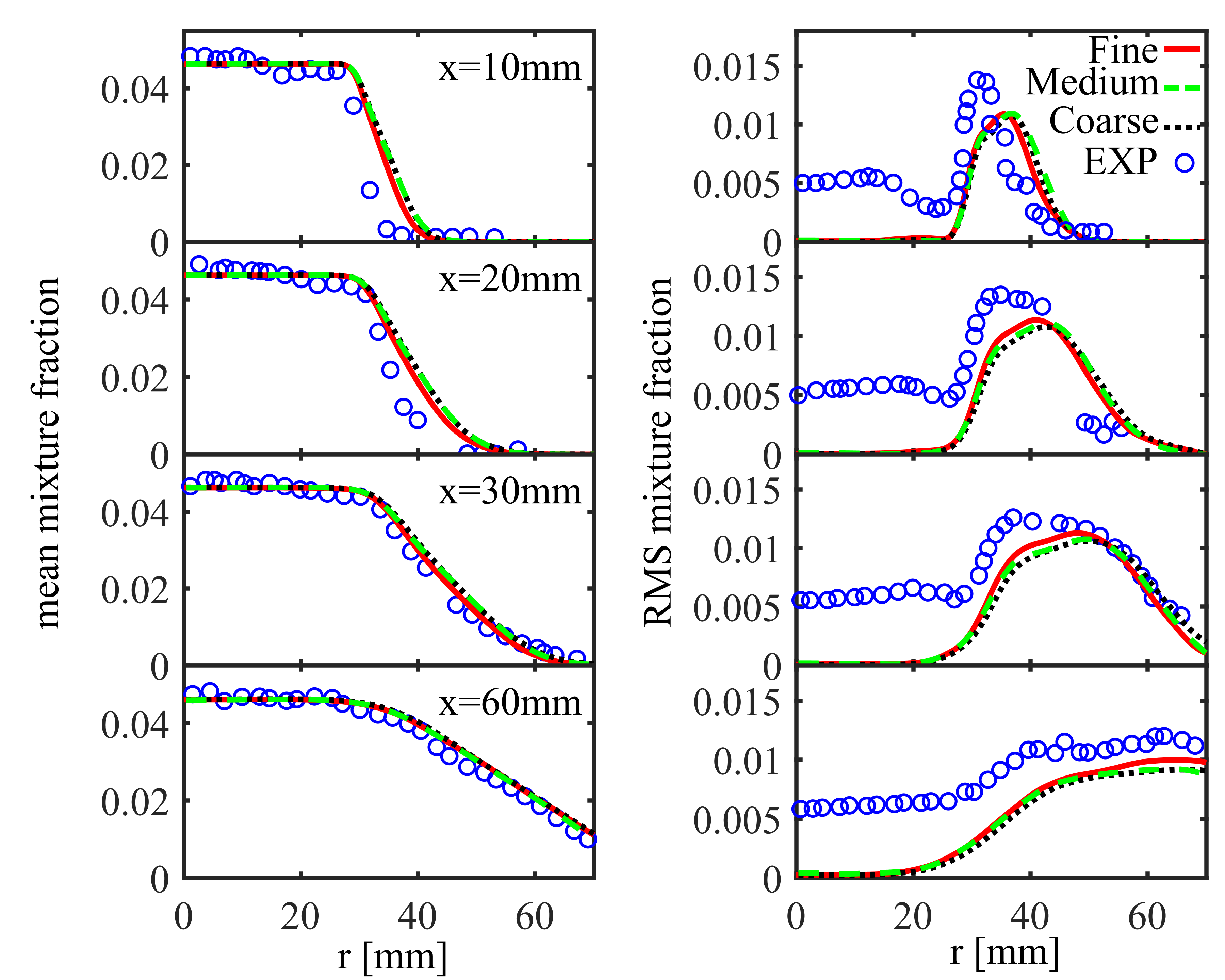}
	\caption{Comparison of radial profiles of mean and fluctuating mixture fractions at different axial positions for the coarse, medium and fine grids.}
    \label{fig:convergence_reac_mix}
\end{figure}

\begin{figure}[!ht]
	\centering
	\includegraphics[width=0.8\textwidth]{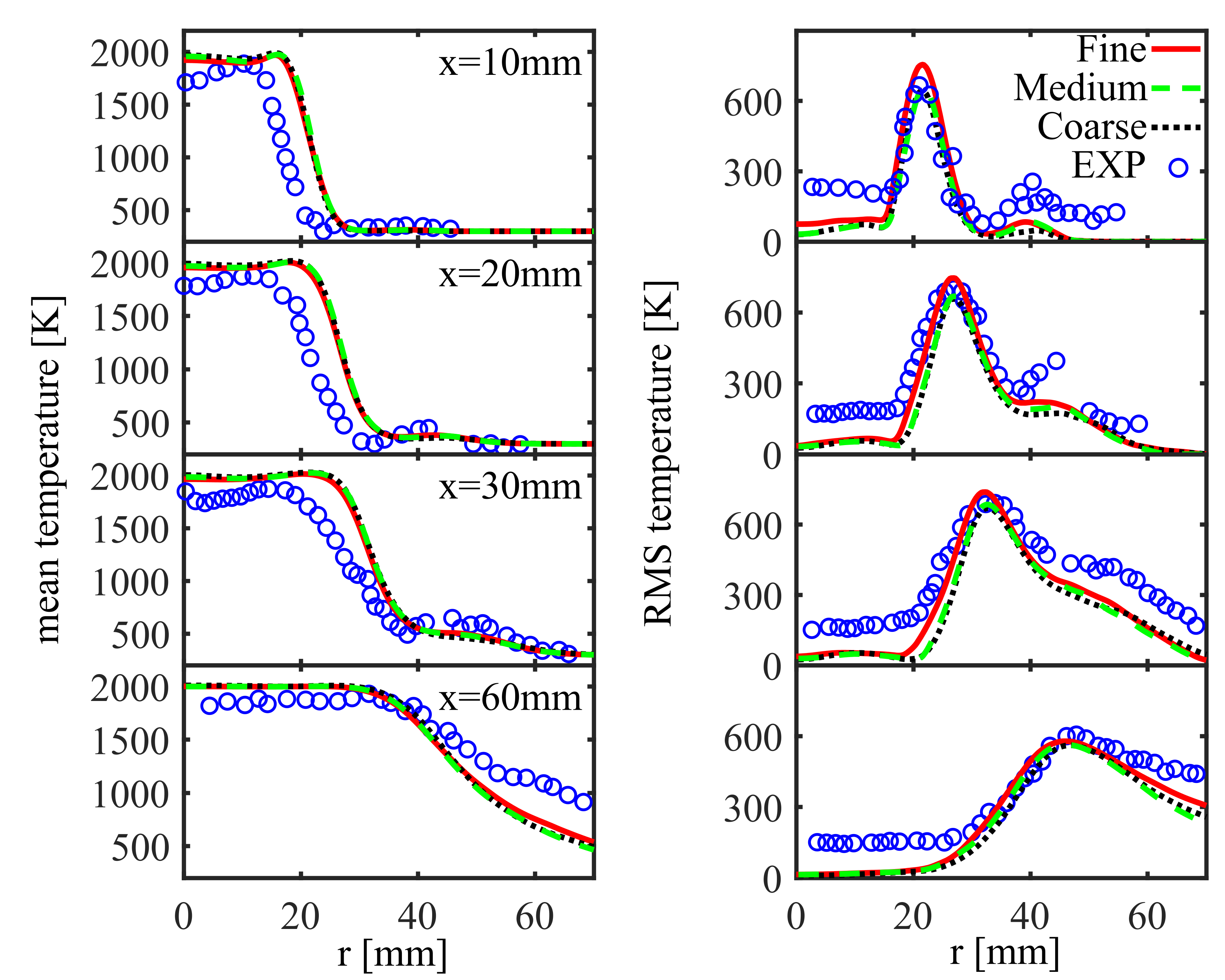}
	\caption{Comparison of radial profiles of mean and fluctuating temperatures at different axial positions for the coarse, medium and fine grids.}
    \label{fig:convergence_reac_temp}
\end{figure}

A grid convergence analysis was conducted for the reactive case, for which three grids with different resolutions were used. 
The coarse and medium grids consist of \(128\times128\times128\) and \(128\times128\times256\) cells, respectively.
The fine grid, which corresponds to the resolution adopted in the present work, consists of \(176\times176\times256\) cells.
Similar to the fine grid, the coarse and medium grids were refined near the reactive region.
Figures~\ref{fig:convergence_reac_axial},~\ref{fig:convergence_reac_radial} and~\ref{fig:convergence_reac_azimuthal} compare the axial, radial, and azimuthal velocities.  
The mean velocity profiles obtained on different grids exhibit a high level of consistency.
The velocity fluctuations are slightly better predicted with the fine grid compared to the coarser grids.
Figures~\ref{fig:convergence_reac_mix} and~\ref{fig:convergence_reac_temp} compare the predicted mixture fraction and temperature statistics.
The mean and fluctuating mixture fraction distributions, as well as the mean temperature, agree well across the different grids.
Slight improvements in the fluctuating temperature are observed with the fine grid.
The overall good agreement among the different grids suggests that the resolution of the fine grid is sufficient for the present work.

\bibliographystyle{elsarticle-num}
\bibliography{cas-refs}
\end{document}